\documentclass[aps,prx,floats,floatfix,twocolumn,superscriptaddress,longbibliography]{revtex4-1}

\usepackage[utf8]{inputenc}
\usepackage{latexsym,amsmath}
\usepackage{amsbsy}
\usepackage[pdftex]{graphicx}
\usepackage{amssymb}
\usepackage{epstopdf}
\usepackage[usenames]{color}
\usepackage{float}
\usepackage{graphicx}
\usepackage{amsmath}
\usepackage[normalem]{ulem}
\usepackage{physics}
\usepackage[colorlinks=true,
bookmarks=false,
linkcolor=red,
urlcolor=red,
citecolor=red,
breaklinks]{hyperref}

\newcommand{\kb}[2]{|#1\rangle\langle#2|}
\newcommand{\ke}[1]{|#1\rangle}
\newcommand{\br}[1]{\langle#1|}

\begin{document}

\title{Experimentally accessible non-separability criteria\\for multipartite entanglement structure detection}

\author{Guillermo Garc\'{i}a-P\'{e}rez}
\affiliation{QTF Centre of Excellence, Department of Physics, Faculty of Science, University of Helsinki, Finland}
\affiliation{Complex Systems Research Group, Department of Mathematics and Statistics,
University of Turku, FI-20014 Turun Yliopisto, Finland}
\affiliation{Algorithmiq Ltd, Linnankatu 55 K 329, 20100 Turku, Finland}
\author{Oskari Kerppo}
\affiliation{QTF Centre of Excellence, Turku Centre for Quantum Physics,
Department of Physics and Astronomy, University of Turku, FI-20014 Turun Yliopisto, Finland}
\author{Matteo A. C. Rossi}
\affiliation{QTF Centre of Excellence, Department of Physics, Faculty of Science, University of Helsinki, Finland}
\affiliation{Algorithmiq Ltd, Linnankatu 55 K 329, 20100 Turku, Finland}
\affiliation{QTF Centre of Excellence, Center for Quantum Engineering, Department of Applied Physics, Aalto University School of Science, FIN-00076 Aalto, Finland}
\author{Sabrina Maniscalco}
\affiliation{QTF Centre of Excellence, Department of Physics, Faculty of Science, University of Helsinki, Finland}
\affiliation{Algorithmiq Ltd, Linnankatu 55 K 329, 20100 Turku, Finland}
\affiliation{QTF Centre of Excellence, Center for Quantum Engineering, Department of Applied Physics, Aalto University School of Science, FIN-00076 Aalto, Finland}

\begin{abstract}
    The description of the complex separability structure of quantum states in terms of partially ordered sets has been recently put forward. In this work, we address the question of how to efficiently determine these structures for unknown states. We propose an experimentally accessible and scalable iterative methodology that identifies, on solid statistical grounds, sufficient conditions for non-separability with respect to certain partitions. In addition, we propose an algorithm to determine the minimal partitions (those that do not admit further splitting) consistent with the experimental observations. We test our methodology experimentally on a 20-qubit IBM quantum computer by inferring the structure of the 4-qubit Smolin and an 8-qubit W states. In the first case, our results reveal that, while the fidelity of the state is low, it nevertheless exhibits the partitioning structure expected from the theory. In the case of the W state, we obtain very disparate results in different runs on the device, which range from non-separable states to very fragmented minimal partitions with little entanglement in the system. Furthermore, our work demonstrates the applicability of informationally complete POVM measurements for practical purposes on current NISQ devices.
\end{abstract}

\maketitle

\section{Introduction}

Quantum correlations are a cornerstone of quantum information theory, entanglement being recognised as the main resource for quantum technologies. The classification and characterisation of non-classical correlations between two parties, including but not limited to entanglement, is well-understood and has been the focus of much literature on quantum theory over the last two decades \cite{Amico2008,HorodeckiRyszard2009Qe,Guhne2009,Modi2012,Adesso2016}.
The extension to multipartite systems, however, faces several challenges. 

The classification of quantum correlations based on local operations and classical communication (LOCC), allowing for their operational definition in the bipartite case, is much more cumbersome in the multipartite case due to both the absence of a maximally entangled reference state \cite{Hebenstreit2016}, and the higher degree of complexity of state transformations \cite{Sauerwein2018}.

Recently, the structure of partial separability and multipartite entanglement has been investigated with the goal of introducing the mathematical formalism to best identify and describe its rich hierarchy \cite{Szalay2012,szalay2015multipartite,Szalay2018,SzalaySzilard2019koea}. For bipartite systems, the fact that there is only one possible partition makes its definition relatively simple. We say that a state $\rho_{AB} \in L \left( \mathcal{H}_A \otimes \mathcal{H}_B \right)$, where $L \left( \mathcal{H} \right)$ represents the space of linear operators in Hilbert space $\mathcal{H}$, is entangled if it cannot be written as $\rho_{AB} = \sum_{i} p_i \rho_A^{(i)} \otimes \rho_B^{(i)}$ with $p_i > 0$ and $\sum_i p_i = 1$, and where $\rho_A^{(i)},  \rho_B^{(i)}$ are quantum states. In the multipartite case, having more than one way of partitioning the system results in a considerably rich \textit{entanglement structure} \cite{szalay2015multipartite} and, consequently, of measures that quantify properties of such structure.

In order to introduce some of these notions in a simple manner, consider a tripartite system composed of subsystems $A$, $B$, and $C$, although the generalisation to $N$-partite systems is straightforward. The absence of entanglement is clear: the state is \textit{fully separable} if it can be written as $\rho_{ABC} = \sum_{i} p_i \rho_A^{(i)} \otimes \rho_B^{(i)} \otimes \rho_C^{(i)}$. If the state is not fully separable, there is some entanglement in the system, but there are several ways in which this can occur.

On the one hand, it may be possible to write the state as $\rho_{ABC} = \rho_{A|BC} \equiv \sum_{i} p_i \rho_A^{(i)} \otimes \rho_{BC}^{(i)}$, or according to another partition, such as $\rho_{AB|C}$ or $\rho_{B|AC}$. The state can be separable with respect to more than one of these partitions. In fact, there exist three-qubit states that are not fully separable ($\rho_{ABC} \neq \rho_{A|B|C}$), and yet admit all three bipartitions, $\rho_{A|BC}$, $\rho_{B|AC}$, and $\rho_{AB|C}$~\cite{DIVINCENZODavidP2003Upbu}. Following Refs.~\cite{szalay2015multipartite,Szalay2018,SzalaySzilard2019koea}, we may call this notion of entanglement, namely with respect to specific set partitions, level-I multipartite entanglement.

On the other hand, it may be possible to express the state as a convex combination of biseparable states, that is, as $\rho_{ABC} = \alpha_1 \rho_{A|BC} + \alpha_2 \rho_{B|AC} + \alpha_3 \rho_{C|AB}$, where $\sum_i \alpha_i = 1, \, \alpha_i \geq 0$, even if no partitions are possible in the sense of level-I entanglement. This different kind of separability, i.e., with respect to multiple set partitions, is referred to as level-II multipartite entanglement. This latter type gives rise to several widely used quantifiers, such as $k$-producibility and $k$-partitionability (also called $k$-separability), as well as to the notion of genuine multipartite entanglement \cite{TamingMultiparticleEntanglement,HorodeckiRyszard2009Qe}, and has widely been investigated experimentally \cite{Friis2018,Lu2018,Saggio2019,Friis2018review}.

Even if level-II entanglement is a very important notion of multipartite entanglement, level-I entanglement can lead to rich and complex structures, the study of which we address in this work. Specifically, we address the following question: given $M$ copies of an $N$-partite quantum system in an unknown state $\rho$, how can we determine its level-I partitioning structure? While we do not provide a complete and general answer to this question, we propose an experimentally accessible and scalable iterative methodology that identifies, on solid statistical grounds, sufficient conditions for non-separability with respect to certain partitions. In addition, we propose an algorithm to determine the minimal partitions (those that do not admit further partitioning) consistent with the experimental observations. 

We test our methodology experimentally on a 20-qubit IBM Quantum computer by inferring the level-I structure of the 4-qubit Smolin \cite{Smolin2001} and an 8-qubit W states. In the first case, our results reveal that, while the fidelity of the state is very low, it nevertheless exhibits the partitioning structure expected from the theory. In the case of the W state, we obtain very disparate results in different runs on the device, which range from one single element in the poset (that is, in which no level-I partitions are possible), in accordance with the theoretical expectations, to very fragmented minimal partitions, revealing little entanglement in the system. To the best of the authors' knowledge, this is the first experimental investigation of level-I entanglement structure. Incidentally, we show the feasibility of informationally complete POVM-based state tomography on current NISQ devices. 

The paper is structured as follows. We first introduce the level-I separability structure of quantum states in Section \ref{sec:sep_struct}, and then present our methodology to determine such structures experimentally in Section \ref{sec:methodology}. In Section \ref{sec:results}, we test the method experimentally on IBM Q quantum computers, and we conclude with a discussion in Section \ref{sec:conclusions}.

\section{Level-I separability structure}\label{sec:sep_struct}

In this work, we focus on level-I multipartite entanglement. While this notion, simpler than level-II, misses details regarding the necessary quantum resources required to prepare the state, it describes a very important aspect: it identifies which subsets of parties do not need to share any quantum resources at all in the preparation. For instance, if $\rho_{ABC} = \rho_{A|BC}$, $A$ and $BC$ can jointly prepare $\rho_{ABC}$ using only local operations and classical communication (LOCC) between them. Moreover, the example explained in the introduction reveals that even for only three parties, the entanglement of the state can be non-trivial with respect to set partitioning.

Mathematically, the level-I entanglement structure of a given state can be associated with a partially ordered set (poset). A poset is a set along with a relation indicating that, given two elements in the set, one precedes the other; in a poset, however, this relation does not apply to all pairs of elements, so its elements can only be considered to be partially ordered. The connection between multipartite entanglement and posets can be established as follows. 

First, suppose that we have a system composed of $N$ parties $\mathcal{S} = \lbrace S_1, \ldots, S_N \rbrace$. Consider a partition $\mathcal{P} = \lbrace P_k \rbrace$ of the system in terms of $|\mathcal{P}|$ non-empty disjoint subsets ($P_k \cap P_{l} = \emptyset \Leftrightarrow k \neq l$) such that $\bigcup_{P_k \in \mathcal{P}} P_k = \mathcal{S}$, as well as another partition $\mathcal{Q} = \lbrace Q_k \rbrace$ that can be obtained by \textit{merging} some of the $P_k$ in $\mathcal{P}$ or, more precisely, such that $\forall P_k \in \mathcal{P}, \exists Q_l \in \mathcal{Q}$ fulfilling $P_k \subseteq Q_l$. Let us call such relation between partitions a \textit{refinement}. We state that $\mathcal{P}$ is a refinement of $\mathcal{Q}$, and write $\mathcal{P} \preceq \mathcal{Q}$; according to this definition, any partition is a refinement of itself. Notice that not all pairs of partitions are related in these terms: given two partitions $\mathcal{A}$ and $\mathcal{B}$, it may be impossible to obtain one of them by merging subsets in the other, that is, $\mathcal{A} \npreceq \mathcal{B}$ and $\mathcal{B} \npreceq \mathcal{A}$ (for example, for $\mathcal{A} = \lbrace \lbrace S_1 \rbrace, \lbrace S_2, S_3 \rbrace \rbrace$ and $\mathcal{B} = \lbrace \lbrace S_2 \rbrace, \lbrace S_1, S_3 \rbrace \rbrace$). Hence, the set of possible partitions with the refinement relation $\preceq$ define a partially ordered set.

Now, let us further suppose that the system is quantum, and that  its joint Hilbert space is $\mathcal{H} = \bigotimes_{i=1}^{N} \mathcal{H}_{S_i}$. If a quantum state $\rho \in L \left( \mathcal{H} \right)$ is separable with respect to partition $\mathcal{P}$,
\begin{equation}\label{eq:sep_p}
\rho = \sum\limits_i p_i \bigotimes \limits_{P_k \in \mathcal{P}} \rho_{P_k}^{(i)}, \quad p_i \geq 0, \quad \sum\limits_i p_i = 1,
\end{equation}
where each of the $\rho_{P_k}^{(i)} \in L \left( \bigotimes_{S_l \in P_k} \mathcal{H}_{S_l} \right)$ is a quantum state, then $\rho$ is also separable with respect to any partition $\mathcal{Q} \succeq \mathcal{P}$,
\begin{equation}\label{eq:sep_q}
\rho = \sum\limits_i p_i \bigotimes \limits_{Q_l \in \mathcal{Q}} \rho_{Q_l}^{(i)}.
\end{equation}
It is sufficient to define 
\begin{equation}\rho_{Q_l}^{(i)} = \bigotimes_{P_k \in \mathcal{P} : P_k \subseteq Q_l} \rho_{P_k}^{(i)}, \quad \forall Q_l \in \mathcal{Q}\end{equation}
to see that this is indeed the case. Therefore, this simple observation reveals that the refinement relation between set partitions is automatically inherited by the level-I separability structure of quantum states. More formally, if we define $\mathcal{F}_{\rho}$ as the set of partitions according to which the state $\rho$ is separable, we can write
\begin{equation}
\label{eq:poset_separability}
\mathcal{P} \preceq \mathcal{Q} \land \mathcal{P} \in \mathcal{F}_\rho \Rightarrow \mathcal{Q} \in \mathcal{F}_\rho.
\end{equation}

The inverse, however, does not necessarily hold, even if $\mathcal{P} \preceq \mathcal{Q}$ is fulfilled. If a state is separable with respect to some partition $\mathcal{Q}$, it may not be separable with respect to a refinement $\mathcal{P} \preceq \mathcal{Q}$ resulting from splitting some sets in the former. As a matter of fact, the only case in which this always occurs is for fully separable states. Given that in such case the partition $\mathcal{R} = \lbrace \lbrace S_1 \rbrace, \ldots, \lbrace S_N \rbrace \rbrace \in \mathcal{F}_\rho$, and that any possible partition $\mathcal{T}$ satisfies $\mathcal{T} \succeq \mathcal{R}$, \eqref{eq:poset_separability} implies that the set of possible partitions and $\mathcal{F}_\rho$ are equal. When some entanglement is present, however, the state of the system is not separable with respect to some partitions. What is more, if a partition $\mathcal{Q}$ is not allowed, neither is any of its refinements; otherwise, one could merge products in the refined decomposition and obtain a valid decomposition in terms of $\mathcal{Q}$. This can also be seen from \eqref{eq:poset_separability}, which implies $\mathcal{Q} \notin \mathcal{F}_\rho \Rightarrow \mathcal{P} \npreceq \mathcal{Q} \lor \mathcal{P} \notin \mathcal{F}_\rho$, that is, refinements of $\mathcal{Q}$ do not belong to $\mathcal{F}_\rho$.

Before we proceed, let us clarify the original motivation of this work. Rather than quantifying the amount of entanglement in a multipartite system, we are interested in determining the poset that characterises the separability structure of an unknown quantum state, provided access to $M$ copies of it. In the next section, we present an iterative methodology that partly achieves this goal.

\section{Methodology}\label{sec:methodology}
In this section, we outline the main points of the method that we propose for assessing level-I entanglement structures, while avoiding technical details when possible. Essentially, our approach exploits the recently proposed partial tomography, in which one reconstructs reduced density operators of the system \cite{CotelrPRL2020,Garcia-Perez2020,bonet-monroig2020nearly}, rather than attempting the often prohibitive full state tomography, along with the following simple observation.

Consider again an $N$-partite system, as well as a subset $\mathcal{U} \subset \mathcal{S}$ of its parties. If the state of the system $\rho$ is separable with respect to some partition $\mathcal{P}$, the reduced state $\rho_\mathcal{U} = \mathrm{Tr}_{\mathcal{S} \setminus \mathcal{U}} \left[ \rho \right]$ is separable with respect to the induced partition on $\mathcal{U}$, defined as $\mathcal{P}_\mathcal{U} = \lbrace P_k \cap \mathcal{U} \vert P_k \in \mathcal{P},  P_k \cap \mathcal{U} \neq \emptyset \rbrace$. This can be shown explicitly,
\begin{equation}\label{eq:reduced_separability}
\rho = \sum\limits_i p_i \bigotimes \limits_{P_k \in \mathcal{P}} \rho_{P_k}^{(i)} \Rightarrow \rho_\mathcal{U} = \sum\limits_i p_i \bigotimes \limits_{P_k \in \mathcal{P}} \mathrm{Tr}_{\bar{P}_k} \left[ \rho_{P_k}^{(i)} \right],
\end{equation}
where $\bar{P}_k = P_k \setminus (P_k \cap \mathcal{U})$. This sets our strategy: we tomographically reconstruct partial states $\rho_\mathcal{U}$ (with small $|\mathcal{U}|$) and determine the entanglement across their bipartitions. According to \eqref{eq:reduced_separability}, every observed forbidden bipartition $\mathcal{B}_\mathcal{U}$ of $\rho_\mathcal{U}$ allows us to eliminate all the $N$-partite partitions $\mathcal{P}$ whose induced partition on $\mathcal{U}$ satisfies $\mathcal{P}_\mathcal{U} \preceq \mathcal{B}_\mathcal{U}$.

We therefore propose to first perform so-called informationally complete (IC) generalised measurements on the system by means of single-party positive operator-valued measures (POVMs). By doing so, the measurement outcomes contain all the information about the quantum state, provided enough copies $M$. In practice, this means that the data enables accurate enough tomographic reconstruction of all the $k$-body reduced density matrices (RDMs) with $k \leq K$ for some $K$. Once all these RDMs have been obtained, we can assess the entanglement across all their bipartitions by classically calculating entanglement measures or monotones. Using the reasoning presented above, any observed entanglement at the reduced level implies a restriction on the global separability structure, and therefore allows us to remove elements from the poset. Importantly, we must be able to determine, given an observed value of some entanglement measure, whether that value is significant or it is a fluctuation resulting from the lack of statistics (number of copies $M$). To that end, we propose to perform a $p$-value test to compare the experimental values with values obtained in classical, noiseless simulations on classically correlated states of the same dimension. Finally, it is convenient to determine the set of minimal partitions in the poset, that is, those that do not admit any further refinement, since the whole poset is fully determined by this set. The overall method can be explained in more detail in terms of five steps:

\begin{itemize}
    \item[1.] \textit{Perform single-party informationally complete (IC) measurements.} To reconstruct the partial states $\rho_\mathcal{U}$, we first need tomographic measurements for the corresponding parties in $\mathcal{U}$. While there are different methods to obtain such data, we propose to use single-party IC-POVMs. The mathematical details of these POVMs, as well as of their implementation for qubits used here, can be found in Appendices \ref{app:ic_povm} and \ref{app:sic_povm}.
    
    The POVM-based strategy is advantageous when considering reduced tomography, especially for the purposes of this work. As described above, the measurement data can be used to reconstruct \textit{all} the $k$-partite density operators in parallel. It is however important to note that the larger $k$ is, the more data is required. More precisely, the number of copies of the state required to infer all the $k$-qubit density operators in an $N$-qubit system with a given statistical confidence scales exponentially in $k$ \cite{Jiang2020optimalfermionto}.
    
    \item[2.] \textit{Reconstruct all the $k$-body reduced density matrices (RDMs) with $k \leq K$.} As described above, the measurement outcomes enable the reconstruction of all the RDMs. To reconstruct the $k$-RDM of a subsystem $\mathcal{U}$ (with $|\mathcal{U}| = k$), we use a likelihood maximisation approach. We first marginalise the outcomes of the $N$-partite local POVM to the subsystems in $\mathcal{U}$ to compute the number of times that each outcome $\mathbf{m}$ has been obtained. Let us denote these frequencies by $f_\mathbf{m}$. We can then proceed to find the state $\rho \in \mathcal{H}_\mathcal{U}$ that maximises the likelihood for these observations to be obtained if one performs the corresponding measurements on $M = \sum_\mathbf{m} f_\mathbf{m}$ copies of the state. This likelihood is given by $L(\rho) = \prod_\mathbf{m} \mathrm{Tr} \left[ \rho \Pi_\mathbf{m}^\mathcal{U} \right]^{f_\mathbf{m}}$, where the operators $\Pi_\mathbf{m}^\mathcal{U}$ are the effects describing the POVM (see Appendix \ref{app:ic_povm} for details). While finding the positive operator that maximises this function is generally non-trivial, this problem has been widely studied and several methods exist \cite{Hradil1997quantumstate,banaszek1999maximum,smolin2012efficient}. We use the diluted maximum-likelihood algorithm from Ref.~\cite{Rehacek2007diluted}. 
    
    \item[3.] \textit{For each $k$-body RDM and every one of its possible bipartitions, calculate an entanglement measure or monotone.} Since the reconstructed states involve only a small number of parties $k \leq K$, we can study their individual entanglement structure. In this work, we consider concurrence \cite{WoottersConcurrence}
    for pairwise ($k=2$) states, which is different from zero if and only if the state is entangled. For $k>2$, we calculate the negativity \cite{VidalWernerNegativity} of the state with respect to each of its bipartitions. This quantity detects entanglement through the negativity of the partial transposition. If the state is separable with respect to a bipartition, the corresponding partial transposition surely yields a positive operator. Hence, negative eigenvalues upon partial transposition necessarily imply entanglement. However, some states result in positive transpositions despite being entangled.
    
    \item[4.] \textit{Filter out statistically irrelevant entanglement observations.} Once the entanglement measures have been computed, one may be tempted to assume that any non-zero value signals the presence of entanglement. However, given that the number of experimental outcomes $M$ is finite, the tomographic reconstruction of the state will not be exact, even in the absence of experimental noise. As a result, we may obtain some non-zero entanglement in our calculation, as a consequence of mere finite statistics, even for separable states. Let us refer to such measured entanglement as \textit{spurious entanglement}.
    
    In order to determine whether the entanglement observed in some subsystem $\mathcal{U}$ is statistically significant or spurious, we propose to use a $p$-value test: given an observed entanglement measure, we assess the probability $p$ for any separable state ---of the same system, across the same bipartition, and reconstructed using the same method and number of copies of the state--- to yield a larger value of the corresponding measure. If this probability is large ($p \geq p_\mathrm{thr}$ for some confidence threshold $p_\mathrm{thr}$ of our choice), the observation is deemed compatible with a fluctuation induced by the lack of statistics, and is therefore discarded. Instead, if $p$ is small, $p < p_\mathrm{thr}$, it is unlikely that the measure is not a consequence of entanglement in the system. In this work, we consider $p_\mathrm{thr} = 0.05$.
    
    To obtain the distribution of spurious entanglement, we simply simulate classically the random outcomes of separable states $\rho_\mathrm{sep}$ that yield high spurious entanglement. More precisely, we use classically correlated states since, according to our numerical experiments, these yield much larger spurious entanglement than other states. In Appendix~\ref{app:statistical_filter} we explain this filtering method in detail. A similar idea, namely using a $p$-value test for entanglement detection, was also proposed in Ref.~\cite{PhysRevA.94.012343}.

    \item[5.] \textit{Remove the incompatible partitions from the poset and identify the minimal partitions.} The statistically relevant non-separability conditions can be now used to remove incompatible partitions from the separability poset. However, this task can be unfeasible for large systems, given that the size of the poset grows very fast with the number of parties. To address this issue, we propose to identify the set of minimal partitions $\mathcal{M}_\rho = \lbrace \mathcal{P} \in \mathcal{F}_\rho \vert \mathcal{Q} \preceq \mathcal{P} \wedge \mathcal{Q} \neq \mathcal{P} \Rightarrow \mathcal{Q} \notin \mathcal{F}_\rho \rbrace$; all the partitions in $\mathcal{F}_\rho$ can be constructed from the elements in $\mathcal{M}_\rho$ through simple merging operations. We present an algorithm enabling to find $\mathcal{M}_\rho$ while avoiding to explore the vast space of all possible set partitions, hence making the problem more approachable. The main working principle of the algorithm, which is detailed in Appendix~\ref{app:minimal_partitions}, is to keep track of the minimal set only, and update it iteratively by considering the sequence of relevant entanglement observations.
    \end{itemize}

\section{Results}\label{sec:results}
We test the proposed methodology experimentally on the 20-qubit IBM Quantum computer \texttt{ibmq\_singapore}. We consider two different scenarios. In the first one, we implement the 4-qubit Smolin state, which exhibits a non-trivial separability structure. In the second case, we prepare an 8-qubit W state, which does not accept any partition whatsoever.

\subsection{Smolin state}

\begin{figure}
    \centering
    \includegraphics[width=0.95\columnwidth]{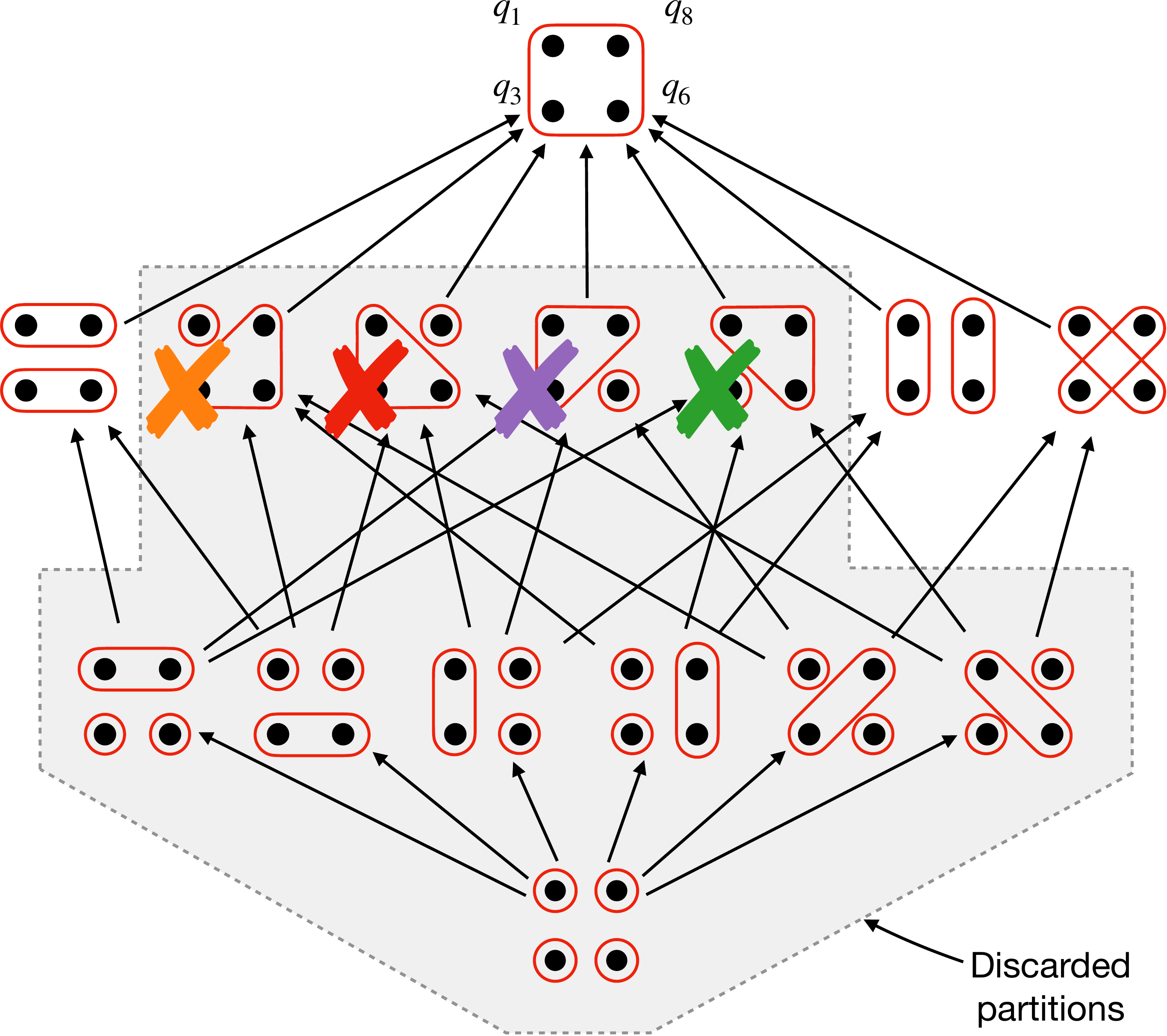}\\
    \vspace{2em}
    \includegraphics[width=0.95\columnwidth]{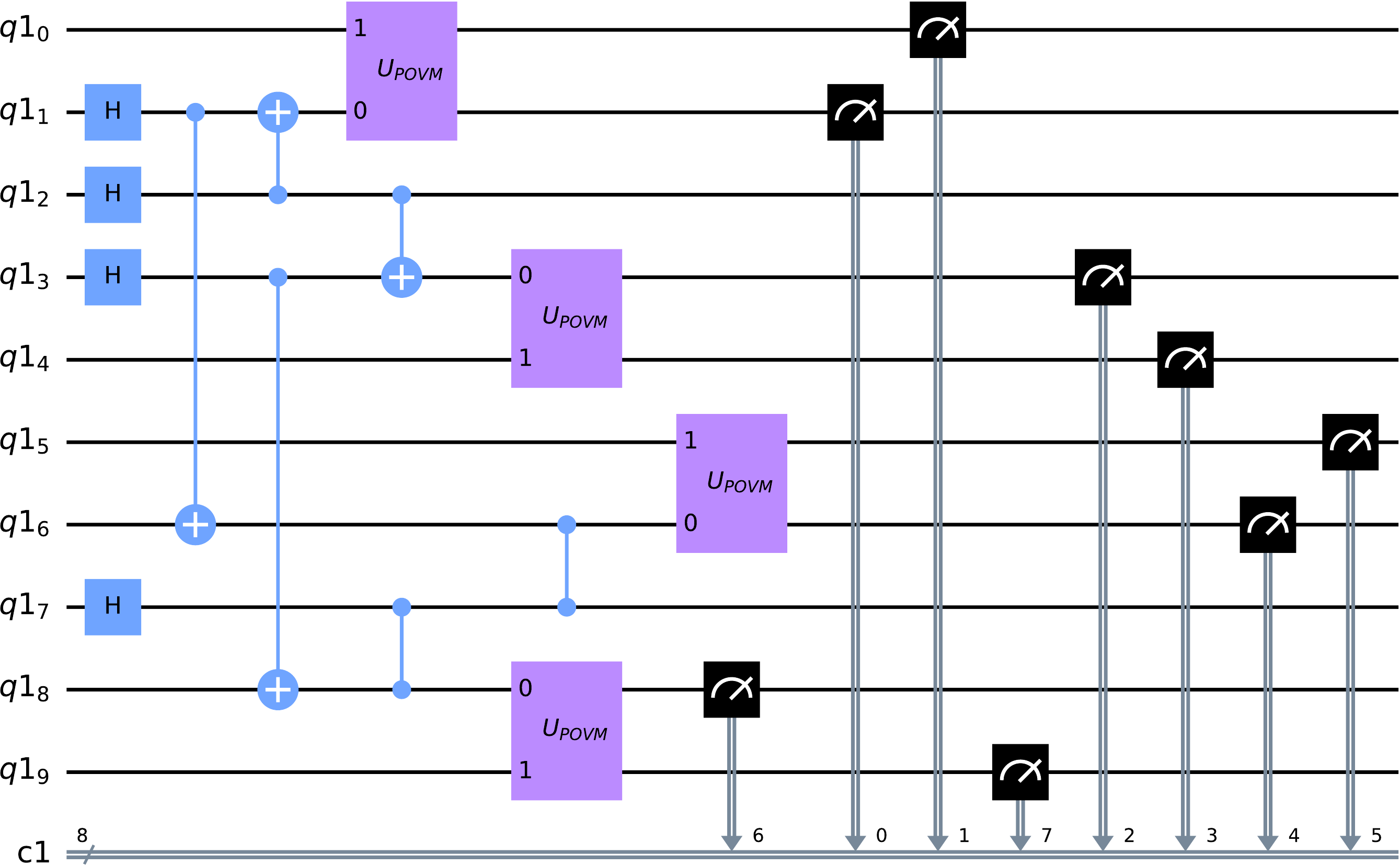}
    \caption{Top: experimental reconstruction of the level-I separability poset of the Smolin state. Partitions are indicated by the red contours. The arrow between two elements indicates the refinement relation, that is, if $\mathcal{A} \preceq \mathcal{B}$, an arrow from $\mathcal{A}$ to $\mathcal{B}$ is drawn. The crossed-out partitions have been observed to yield negativity in the tomographic reconstruction. The colour of the cross identifies the observed negativity value in Fig.~\ref{fig:smolin_results}. Given that these partitions are not permitted, none of their refinements are. Therefore, the only partitions allowed, according to the observations, are the ones not covered by the grey area. Bottom: circuit decomposition of the Smolin state preparation circuit with the POVM measurement. The qubit assignment takes into account the connectivity of the device (see Appendix \ref{app:smolin_state} for details).}
    \label{fig:smolin_poset_circuit}
\end{figure} 

\begin{figure*}
    \centering
    \includegraphics[width=\textwidth]{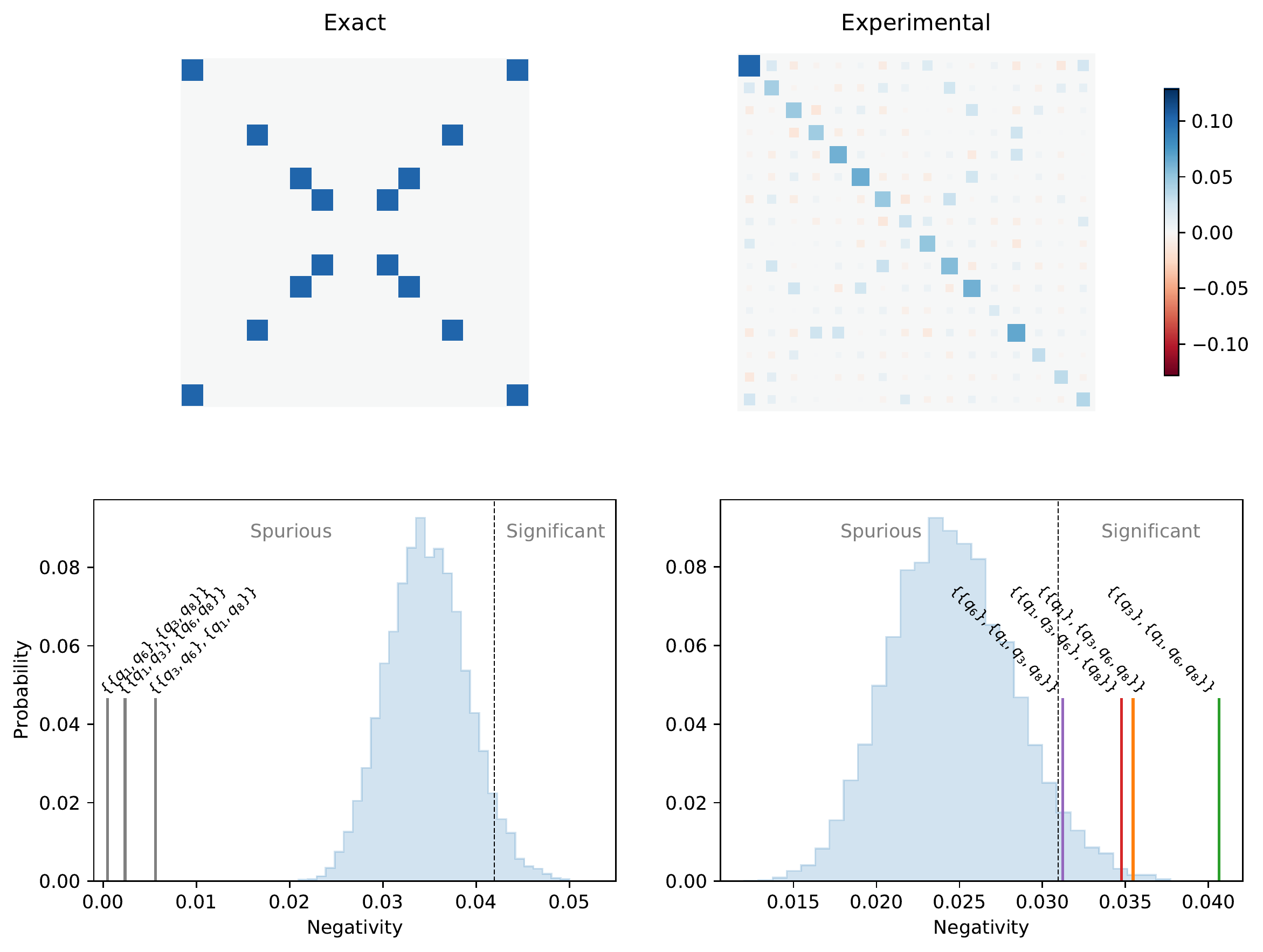}
    \caption{Top: exact (left) and experimental, tomographically reconstructed (right) real part of the density matrix of the Smolin state. The fidelity of the reconstructed state with the exact state is approximately 0.64. Bottom: experimental negativity with respect to $(2, 2)$ (left) and $(1,3)$ (right) bipartitions (solid vertical lines), compared to the histogram of spurious entanglement obtained from highly correlated separable states. The dashed vertical line marks the separation between the region of spurious entanglement (left region) and statistically significant entanglement (right region). The $(2,2)$ bipartitions have a non-zero, but statistically insignificant negativity, whereas the $(1,3)$ have a negativity that, although much lower than the expected one, is above the threshold of spurious entanglement. This allows us to remove these partitions in the poset of Fig.~\ref{fig:smolin_poset_circuit} and find the expected entanglement structure of the Smolin state.}
    \label{fig:smolin_results}
\end{figure*} 

The 4-qubit Smolin state is an interesting case study for our methodology. It is defined as a statistical mixture of products of Bell states,
\begin{equation}\label{eq:smolin_state}
\rho = \frac{1}{4} \sum\limits_{i=0}^{3} \ke{ \Psi^{(12)}_{i} } \br{ \Psi^{(12)}_{i} } \otimes \ke{ \Psi^{(34)}_{i} }  \br{ \Psi^{(34)}_{i} },
\end{equation}
where $\lbrace \ke{ \Psi^{(12)}_{0} }, \ldots, \ke{ \Psi^{(12)}_{3} } \rbrace$ represent the four Bell states of the qubits in their superscript. This state is separable with respect to any bipartition $\mathcal{P} = \lbrace P_1, P_2 \rbrace$ such that $\vert P_1 \vert = \vert P_2 \vert = 2$. Using the notation $(\vert P_1 \vert, \vert P_2 \vert)$ to characterise the subset sizes, we may refer to these partitions as $(2, 2)$-bipartitions. On the other hand, it is negative with respect to the partial transposition of $P_1$ (or $P_2$) if $\vert P_1 \vert = 1$. In other words, it is not separable with respect to $(1, 3)$-bipartitions such as $\rho_{1 \vert 234}$, $\rho_{2 \vert 134}$, etc. In this case, the size of the system is small enough for us to represent the whole level-I entanglement poset of the state (see Fig.~\ref{fig:smolin_poset_circuit} (top)). 

It may result illustrative to discuss the separability structure of the state in Eq.~\eqref{eq:smolin_state} in more detail. First, notice that, given that any further partition of a $(2, 2)$-bipartition leads to a refinement of a $(1, 3)$-bipartition, $\nexists \mathcal{Q} \in \mathcal{F}_\rho$ such that $\mathcal{Q} \preceq \mathcal{P} \wedge \mathcal{Q} \neq \mathcal{P}$ if $\mathcal{P}$ is a $(2, 2)$-bipartition. Moreover, every allowed partition is either a $(2, 2)$-bipartition or the trivial partition in which all parties belong to the same set so, for this state, $\mathcal{M}_\rho$ is equal to the set of $(2, 2)$-bipartitions.

Now, let us use this state as an example for an experimental implementation of our methodology. Since we are using a gate-based quantum computer, we need to devise a circuit to prepare the state. However, the state in Eq.~\eqref{eq:smolin_state} is not pure, so we must use ancillary qubits (in particular, we need at least two in order to create a rank-4 state). The strategy to do so is rather simple: first, prepare two copies of the Bell state $\ke{\Psi^{(ij)}_{0}} = (\ke{00} + \ke{11}) / \sqrt{2}$ by using a Hadamard-CNOT sequence on two pairs of qubits. This Bell state can be transformed into any other Bell state by either applying an $X$ gate on one qubit, a $Z$ gate on the other, or both. Second, prepare two other (auxiliary) qubits in an equal superposition of all possible computational basis states (by using two Hadamard gates). Finally, each of these auxiliary qubits acts as the control of the corresponding controlled-$X$ or -$Z$ gates that transform the Bell states (see Fig.~\ref{fig:smolin_poset_circuit} (bottom)). The resulting state is thus $\sum_{i=0}^3 \ke{i} \otimes \ke{ \Psi^{(12)}_{i} } \otimes \ke{ \Psi^{(34)}_{i} } / 2$, where $\ke{i}$ refers to the state of the auxiliary two-qubit system. Tracing out (disregarding) the auxiliary qubits yields Eq.~\eqref{eq:smolin_state}. To conclude, we must add the measurement circuit implementing the SIC-POVM on each of the four qubits, which requires four additional qubits in the ground state (see Appendix~\ref{app:sic_povm}). In total, the experimental setup involves 10 qubits. In practice, one must also take into account the connectivity of the device when deciding the role played by every physical qubit. See Appendix~\ref{app:smolin_state} for a description of the experimental details.

We ran the resulting final circuit $20$ times with the maximum number of shots allowed by the IBM Quantum devices, $8192$. Hence, in total, we used $163840$ copies of the state in our experiments. In accordance with our procedure, we exploit the informationally complete data to reconstruct all the $k$-qubit states with $k \leq K$. In this case, we set $K = 4$ so, in fact, we are performing full state tomography, given the small system size. The states with $k < 4$ are highly mixed (so the noisy data leads to high-fidelity reconstructions, with fidelities above 0.96; yet, they are of little use to us, given that they are separable). Instead, the full 4-qubit state reconstruction yields interesting results.

On the one hand, we note that the fidelity of the reconstructed state is rather low (approximately 0.64). In Fig.~\ref{fig:smolin_results} (top), we show the reconstructed density matrix next to the theoretical one. While some of the structure in the matrix seems to be partially reproduced, the differences are notable. The assessment of the negativity with respect to bipartitions reveals very low values, both for $(2, 2)$- (for which they should be zero) and $(1, 3)$-bipartitions (for which they should be high). However, there are significant statistical differences when it comes to the corresponding precise values. In Fig.~\ref{fig:smolin_results} (bottom), we depict these values along with the distribution of spurious entanglement obtained from highly correlated separable states. Remarkably, the values corresponding to the $(2, 2)$-bipartitions lie well within the bounds of spurious entanglement for separable states, whereas those of $(1, 3)$-bipartitions do not. This statistical analysis serves the purpose of providing a context to the numerical values obtained, so that experimental entanglement measures can be deemed ``high'' or ``low'' with respect to some meaningful benchmark.

These results therefore reveal that, despite the considerable levels of noise in the quantum device, which lead to low state fidelity, the physical operations within the processor lead to a state with the expected entanglement structure. This is indeed consistent with the fact that, even though the obtained fidelity, $0.64$, seems very low, it is relatively large when compared to randomly chosen states (see Appendix \ref{app:fidelity}), which indicates that the reconstructed state retains some similarity to the theoretically expected one.

\subsection{W state}\label{sec:w_state}
\begin{figure*}[t]
    \centering
        \raisebox{-0.5\height}{\includegraphics[width=\columnwidth]{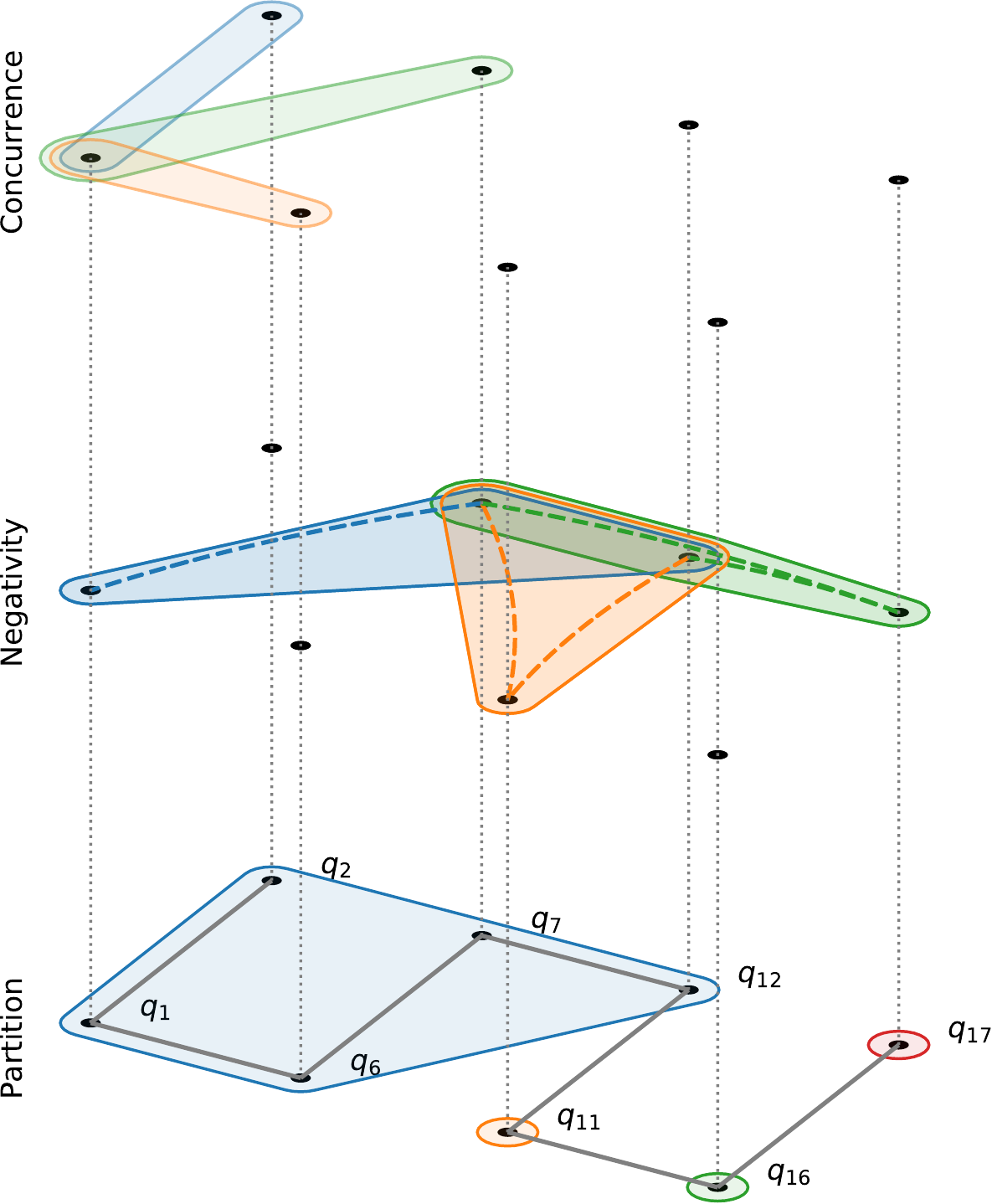}}\hfill
        \raisebox{-0.5\height}{\includegraphics[width=\columnwidth]{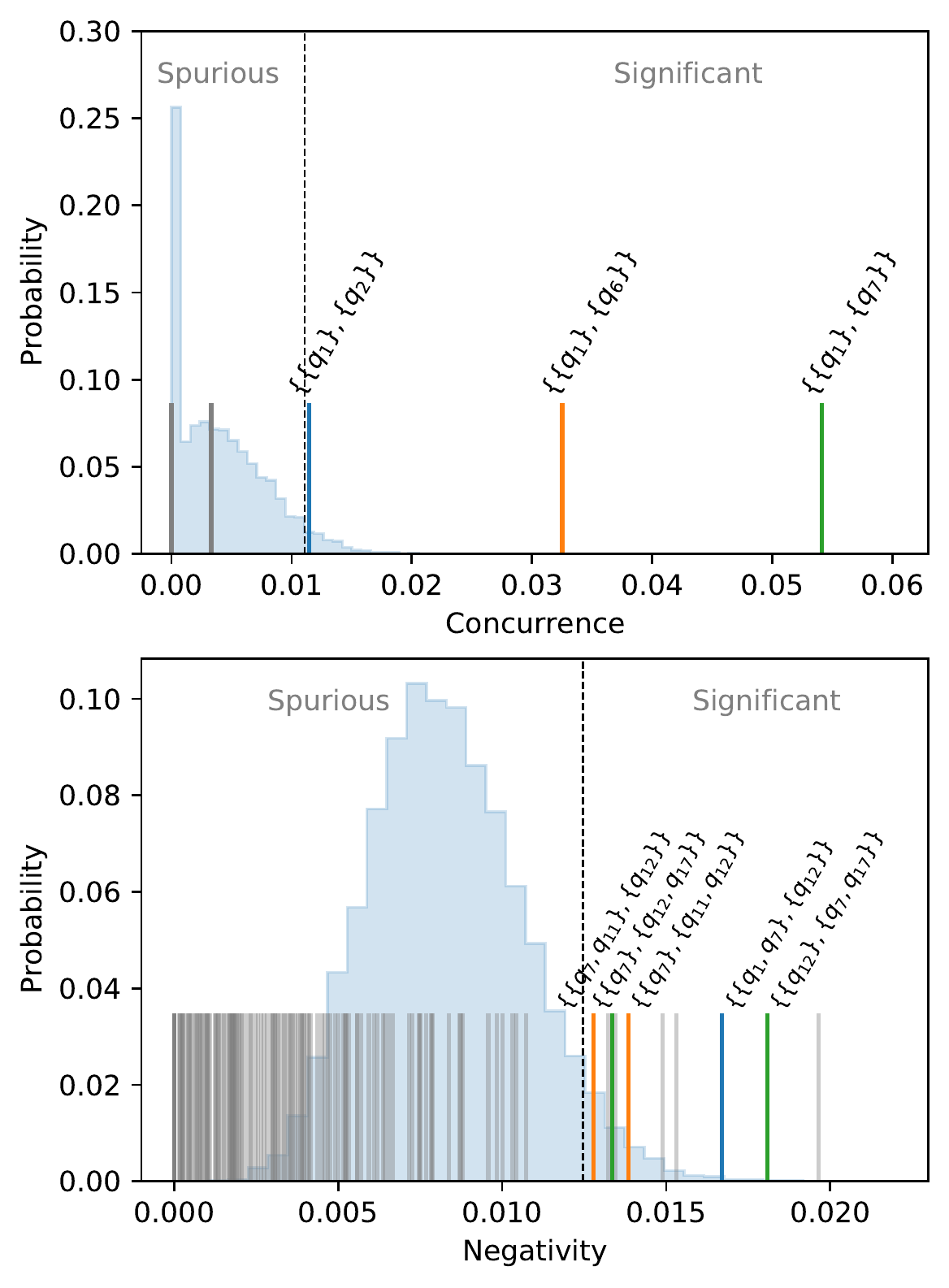}}
    \caption{Left: an example multiplex plot of the experimentally reconstructed entanglement structure for the input W state. From top to bottom, the three layers depict the statistically relevant concurrence, the three-qubit negativity (dashed links join qubits that belong to a two-qubit set), and the only minimal partition in $\mathcal{M}_\rho$ in this case, from which all other permitted partitions can be obtained by merging sets. Right: the solid vertical lines correspond to experimental values for the concurrence of qubit pairs (top) and negativity of $(1, 2)$ partitions (bottom), compared to distributions of the same quantities for highly correlated separable states. The coloured lines are the significantly entangled pairs/partitions that are drawn in the multiplex plot.}
    \label{fig:w_state_results}
\end{figure*}
We now turn our attention to a state that is simpler than the Smolin state from the point of view of its level-I entanglement structure. The $N$-qubit W state $\ke{\mathrm{W}_N}$ is defined as an equally balanced superposition of single-excitation basis states, that is,
\begin{equation}\label{eq:W_state}
    \ke{\mathrm{W}_N} = \frac{1}{\sqrt{N}} \left( \ke{0\ldots01} + \ke{0\ldots10} + \cdots + \ke{1\ldots00} \right).
\end{equation}
In this state, every two-qubit reduced state is entangled, with a concurrence equal to $2/N$. Hence, given few copies of the W state, our approach would start with the measurement of these concurrences, which would immediately lead to the conclusion that no partitions are possible whatsoever. However, the reason why this state can be illustrative for this work is that, in an actual implementation, much of this entanglement can be lost. Hence, it is nevertheless interesting to assess to what extent, and with what structure, qubits become entangled in the physical processor when aiming at the preparation of the state in Eq.~\eqref{eq:W_state}.

Moreover, another advantage of W states is that there exist efficient algorithms for their preparation on gate-based processors \cite{Cruz2019}, although the efficiency of these approaches is largely reduced by the restrictions imposed by the connectivity of the device. In short, the working principle of the state preparation algorithm is as follows. We first initialise two qubits in an entangled state of the form $\alpha \ke{01} + \sqrt{1 - \alpha^2} \ke{10}$. Next, we correlate the rest of the qubits sequentially by applying two-qubit gates that preserve the number of ones in the state. In this work, we consider the 8-qubit W state $\ke{\mathrm{W}_8}$. In Appendix~\ref{app:W_state}, we explain the procedure in more detail. In the Supplementary Material (SM), we show the 16-qubit circuit run on the IBM Quantum computer \texttt{ibmq\_singapore} resulting in the state preparation including the POVM using auxiliary qubits.

As with the Smolin state, we ran the circuit $163840$ times by submitting $20$ $8192$-shot jobs. We also repeated the experiments several times, obtaining very disparate results. The IBM Quantum devices are calibrated on a daily basis, and the same circuit run on different days can yield considerably different outcomes.

Given that the poset for 8 qubits is too large to be depicted, we introduce a different graphical representation, namely in terms of multiplex hypergraph plots in which we only draw the minimal partitions $\mathcal{M}_\rho$ (from which all other permitted partitions follow trivially by merging sets), along with a minimal subset of statistically relevant entanglement observations leading to those partitions. The method to find $\mathcal{M}_\rho$ is explained in Appendix \ref{app:minimal_partitions}. In Fig.~\ref{fig:w_state_results} (left), we show an example of such representation of a minimal partition $\mathcal{P} \in \mathcal{M}_\rho$, which we now detail.

We start by drawing the qubits according to some layout on the plane. In the case of Fig.~\ref{fig:w_state_results} (left, bottom diagram), the layout matches that of the device (cf. Fig.~\ref{fig:singapore_layout}). Moreover, we have complemented the representation by indicating the connectivity of the device. A partition on that set of qubits can then be represented simply by covering the qubits with disjoint areas, like the coloured ones depicted in the figure. The different entanglement observations can also be drawn on the layout in a similar manner. However, in order to prevent too much information overlapping in the same plot, we represent the entanglement observations in different layers, stacked vertically.

The entanglement observations are represented as follows. First, we consider all the values of concurrence obtained from the experiment, and we filter out those that are not deemed statistically relevant according to the $p$-value test. This procedure is shown in Fig.~\ref{fig:w_state_results} (top right). Each relevant value of concurrence is drawn in the concurrence layer as an area enclosing the two entangled qubits. In principle, one could do the same thing for the three-qubit negativity. However, we can simplify the representation somewhat by noticing that some of the statistically relevant negativity values are implied by the observed concurrences. For instance, suppose that the qubits $S_1$ and $S_2$ have relevant concurrence, so the partition $\lbrace \lbrace S_1 \rbrace, \lbrace S_2 \rbrace \rbrace$ is not permitted. Then, a negativity in the partition $\lbrace \lbrace S_1 \rbrace, \lbrace S_2, S_3 \rbrace \rbrace$ does not add any information to the separability of the state, as it is implied by the former. Thus, we can draw only the statistically relevant negativity observations that are not implied by the concurrence; these are the observations highlighted in colour in Fig.~\ref{fig:w_state_results} (bottom right). The negativity observations are represented by a coloured area covering the three qubits involved. To indicate to which specific bipartition of the three qubits the area refers, we add a dashed line between the two qubits in the same subset. For instance, in Fig.~\ref{fig:w_state_results}, for the negativity of the bipartition $\lbrace \lbrace q_{12} \rbrace, \lbrace q_1, q_7 \rbrace \rbrace$ (blue), we add a dashed line between qubits $q_1$ and $q_7$.

This representation allows us to see at a glance the entanglement structure of the state along with the relevant experimental observations that lead to conclude that further partitioning of the state is not possible. In the example presented in Fig.~\ref{fig:w_state_results}, it is easy to see that the cluster formed by qubits $q_1$, $q_2$, $q_6$, and $q_7$ cannot be further split because of the observed concurrences. Notice that the value of concurrence between $q_1$ and $q_2$ is very close to the threshold of the statistical filter, which means that if we decide to be more restrictive regarding what constitutes a statistically relevant observation and reduce the value of $p$ in the $p$-value test, this value will no longer be considered. Interestingly, the most significant value is obtained for qubits $q_1$ and $q_7$, which are not connected in the device.

Another interesting conclusion that can be drawn from these observations is that qubit $q_{12}$ must also belong to the same cluster, despite not exhibiting any relevant concurrence with any other qubit. Instead, this is implied by the negativity, arguably in two distinct ways. On the one hand, it is straightforwardly implied by the $\lbrace \lbrace q_{12} \rbrace, \lbrace q_1, q_7 \rbrace \rbrace$ observation (blue area). On the other hand, it is also implied in a less straightforward manner by the double negativity observations with qubits $q_{11}$ and $q_{17}$ (orange and green areas, respectively). Indeed, even if the $\lbrace \lbrace q_{12} \rbrace, \lbrace q_1, q_7 \rbrace \rbrace$ negativity had not been observed, we would conclude that $q_{12}$ belongs to the cluster: since $\lbrace \lbrace q_{12} \rbrace, \lbrace q_{11}, q_7 \rbrace \rbrace$ and $\lbrace \lbrace q_{7} \rbrace, \lbrace q_{11}, q_{12} \rbrace \rbrace$ are both forbidden, the partition $\lbrace \lbrace q_7 \rbrace, \lbrace q_{12} \rbrace \rbrace$ must be forbidden too. A similar analysis follows for the green area. In other words, in a similar situation in which the $\lbrace \lbrace q_{12} \rbrace, \lbrace q_1, q_7 \rbrace \rbrace$ negativity had not been observed, we would still be able to conclude that qubit $q_{12}$ must be entangled with the larger cluster, but this conclusion would rely on observations involving qubits outside the cluster. This highlights the inherent complexity in determining the entanglement structure of multipartite states.

The previous analysis describes the results obtained for one realisation of the experiment. In the SM we also include the results for other experimental runs, in Fig. \ref{fig:w_multiplexes_left} and Fig. \ref{fig:w_multiplexes_right}. Overall, the results are very disparate. While in some cases little entanglement is observed, in other realisations, our approach enables us to conclude that the state is not separable with respect to any partition whatsoever, or according to very few. This is remarkable considering the complexity of the quantum circuit executed (see Fig.~\ref{fig:w_circuit}), which involves 16 qubits in total.

\section{Conclusions and outlook}\label{sec:conclusions}
Quantum entanglement in multipartite systems can generally lead to complex structures, which have been thoroughly studied and characterised in the literature. While many approaches have addressed the issue of entanglement certification (that is, validating the presence of entanglement assuming previous knowledge about the system's state), fewer works address its detection for unknown states. In this work, we provide a methodology based on reduced tomography that enables to partially uncover the underlying separability structure of unknown quantum states. Moreover, the method is scalable, in the sense that it does not rely on full state tomography nor on any other technique requiring exponentially scaling resources, as well as iterative; one can further discard partitions from the poset by adding more experimental data and reconstructing larger reduced states. We also provide a classical algorithm that enables to identify the minimal partitions compatible with the observations for moderate system sizes.

We have also tested our approach on a real 20-qubit quantum computer from the IBM Quantum service, with experiments involving 10- and 16-qubit circuits of considerable depth. Our results show that the method is capable of unveiling the correct entanglement structure, that is, the separability poset, despite the low state fidelity due to experimental noise. The statistical validation of the entanglement observation plays an important role in this, as it allows us to clearly discern between statistically relevant and irrelevant observations, even when the observed entanglement is weak. Incidentally, our work shows that POVM-based tomography is feasible in current NISQ devices, which in turn can enable numerous applications.

This article focuses on so-called level-I entanglement, that is, on the representability of the state in terms of a convex combination of states separable with respect to a fixed partition. While this kind of entanglement has an important physical interpretation, namely, it determines which subsets of parties do not need quantum resources to prepare the state, one is often interested in separability with respect to non-fixed partitions. Thus, we plan to extend the ideas outlined in this paper to the more complex scenario of level-II entanglement in the future.

\section*{Acknowledgements}
We acknowledge the use of IBM Quantum services for this work. The views expressed are those of the authors, and do not reflect the official policy or position of IBM or the IBM Quantum team. G.G.-P., O.K., M.A.C.R., and S.M. acknowledge financial support from the Academy of Finland via the Centre of Excellence program (Project no. 312058). S.M.~and G.G.-P. acknowledge support from the emmy.network foundation under the aegis of the Fondation de Luxembourg. G.G.-P. acknowledges support from the Academy of Finland via the Postdoctoral Researcher program (Project no. 341985). O.K. acknowledges financial support from the Turku University Foundation.

\appendix

\section{Informationally complete measurements}\label{app:ic_povm}
A POVM on one of the system parties $i$ is defined in terms of a set of positive semidefinite operators $\lbrace \Pi_m^{(i)} \rbrace \subset L (\mathcal{H}_{S_i})$ fulfilling $\sum_m \Pi_m^{(i)} = \mathbb{I}$ and $\Pi_m^{(i)} \geq 0, \, \forall m$. Given these operators, which are often called \textit{effects}, the probability for a system in state $\rho^{(i)}$ to yield outcome $m$ upon measurement is $\mathrm{Tr} [\Pi_m^{(i)} \rho^{(i)}]$. If the set contains a subset of $\dim (\mathcal{H}_{S_i})^2$ linearly independent effects, they form a basis of $L (\mathcal{H}_{S_i})$, and the state of the system can be reconstructed from the corresponding outcome statistics: the POVM is said to be informationally complete (IC).

Importantly, if one such generalised measurement can be implemented on each party in the subset $\mathcal{U}$, their joint measurement is described by the effects $\Pi_\mathbf{m}^\mathcal{U} \equiv \bigotimes_{i \in \mathcal{U}} \Pi_{m_i}^{(i)}$, where $\mathbf{m}$ represents the collection of single-party outcomes for all parties in $\mathcal{U}$ (that is, a list of the corresponding $m_i$ appearing in the decomposition of the effect). The resulting set of joint effects, $\lbrace \Pi_\mathbf{m}^\mathcal{U} \rbrace$, is also an IC-POVM in $\mathcal{H}_\mathcal{U} =  \bigotimes_{i \in \mathcal{U}} \mathcal{H}_{S_i}$. In other words, these local measurements, if implemented on every party in the system, yield data enabling full state tomography, as well as reduced tomography of any subsystem. Needless to say, full state tomography becomes impractical even for relatively small systems, since it generally requires a large number of measurements and unattainable classical resources to encode the corresponding density operator of the system.

\section{Circuit implementations}
\begin{figure}
    \includegraphics[width=0.8\columnwidth]{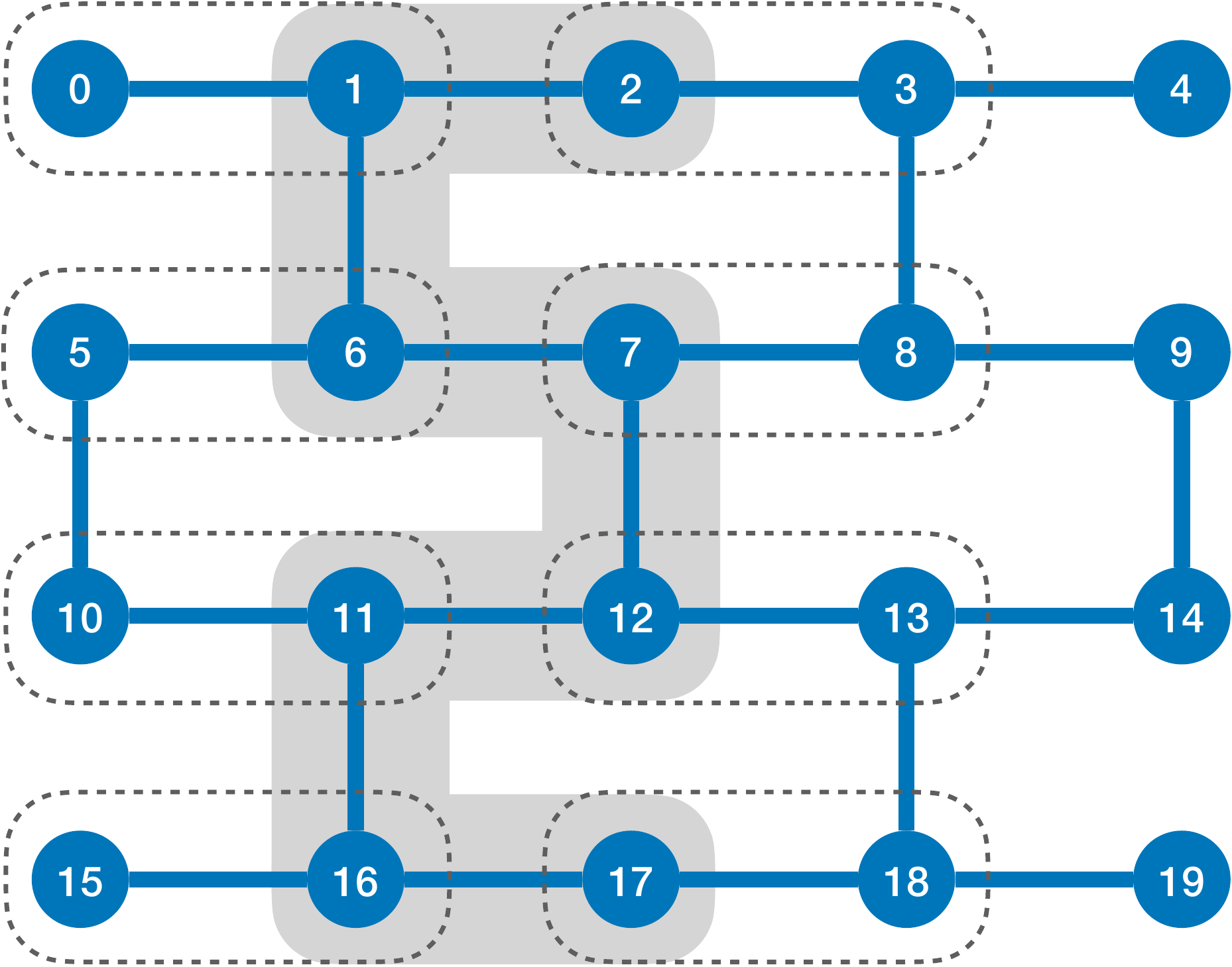}
    \caption{Layout of the \texttt{ibmq\_singapore} device. Blue lines represent physically implementable CNOT gates. The gray shaded area shows the qubits used to prepare the W state, while the dashed lines show the pairing with ancillary qubits in order to perform the POVM measurement.}
    \label{fig:singapore_layout}
\end{figure}

This section contains experimental considerations regarding the circuit implementation of the SIC-POVM, as well as of the two studied states, on the \texttt{ibmq\_singapore} quantum computer. The connectivity of the device, depicted in Fig.~\ref{fig:singapore_layout}, must be taken into account when designing the circuits. In particular, if a two-qubit operation between two disconnected qubits in the device is required, the compiler adds a sequence of SWAP gates in order to transfer the state of the qubits to neighbouring ones, which increases the effect of noise in the experiment notably.

\subsection{SIC-POVM}\label{app:sic_povm}
We apply our methodology to $N$-qubit systems. We thus implement an IC-POVM on each qubit by means of a single-qubit dilation. That is, to every qubit, we associate an extra ancillary qubit in some known state ($\ket{0}$). We then apply a joint unitary to both qubits and consequently measure them. The four possible outcomes are associated to four effects, hence defining a POVM. If the four effects are linearly independent, the POVM is IC. In this work, we consider a specific POVM whose effects are given by $\lbrace \Pi_i = \tilde{\Pi}_i / 2 \rbrace$, where $\tilde{\Pi}_i = \ket{\tilde{\pi}_i} \bra{\tilde{\pi}_i}, \ket{\tilde{\pi}_0} = \ket{0}, \, \ket{\tilde{\pi}_k} = (\ket{0} + \sqrt{2} e^{i 2 \pi (k-1) / 3} \ket{1}) / \sqrt{3}, \, k \in [1, 3]$ are rank-one projectors. These projectors form a regular tetrahedron in the Bloch sphere, so the POVM is considered to be a symmetric IC-POVM (SIC-POVM).

\begin{figure}
    \centering
    \raisebox{-.5\height}{\includegraphics[height=3em]{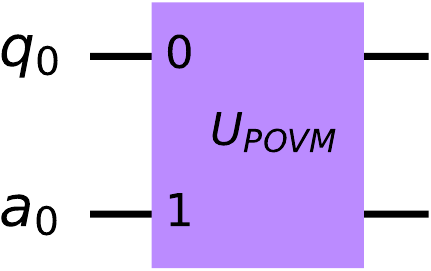}} = \raisebox{-.5\height}{\includegraphics[height=3em]{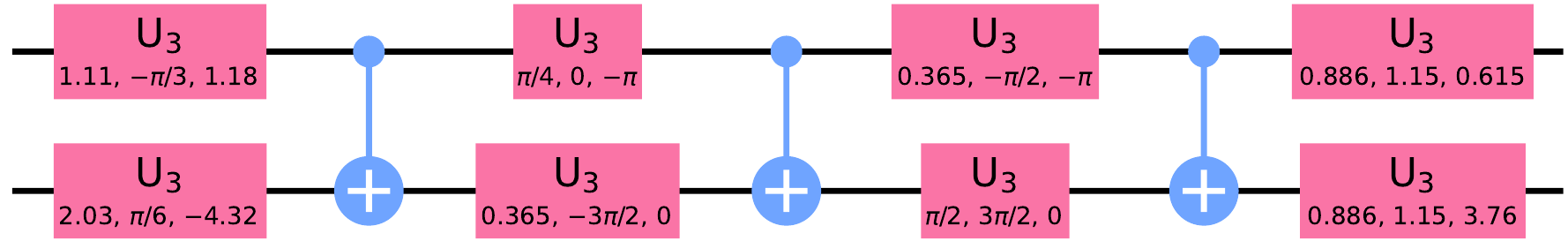}}
    \caption{Gate decomposition of the SIC-POVM unitary on $q_0$ using $a_0$ as auxiliary qubit, using the basis gates available on the IBM Quantum hardware.}
    \label{fig:povm_circuit}
\end{figure}

The correspondence between the qubit-ancilla unitary and the corresponding effects can be easily obtained (see e.g.~Ref.~\cite{garciaperez2021learning} for details), and the specific unitary used for the implementation of the SIC-POVM is included in the code accompanying this publication \cite{github}. The two-qubit unitary can be decomposed as a sequence of single-qubit gates and CNOTs, as shown in Fig.~\ref{fig:povm_circuit}.

\subsection{Smolin state}\label{app:smolin_state}
As explained in the main text, the preparation of the Smolin state requires preparing two Bell states and then applying unitaries to them controlled by two additional qubits. Given the connectivity of the layout in Fig.~\ref{fig:singapore_layout}, we choose two pairs of qubits, $(q_1, q_6)$ and $(q_3, q_8)$, to prepare the initial Bell states. Qubits $q_2$ and $q_7$ can therefore interact with qubits $q_1, q_3$, and $q_6, q_8$, respectively. The two qubits are initially prepared in the $\ket{+}$ state, and then controlled operations, controlled-$X$ and controlled-$Z$ respectively, on their neighbours are applied, resulting in the Smolin state of qubits $q_1, q_3, q_6$, and $q_8$. Qubits $q_0, q_4, q_5$, and $q_9$ can be used as ancillae for the POVM measurement. The corresponding circuit implementation is depicted in Fig.~\ref{fig:smolin_results}. The compiled circuit, decomposed in terms of the native gates of the device, is depicted in Fig.~\ref{fig:smolin_circuit} of the SM and provided with the accompanying code \cite{github}.

\subsection{W state}\label{app:W_state}
As explained in the main text, our strategy to prepare a W state on the quantum processor is similar to the one proposed in Ref.~\cite{Cruz2019}. In principle, the state could be prepared in linear time by applying single-excitation-preserving gates sequentially along a chain of qubits. However, what the authors of the aforementioned paper propose is to parallelise the sequence of gates as much as possible. If the topology of the device allows it, the state can be prepared using a circuit of depth logarithmic in the system size. With this in mind, we proceed in the following way. First, we entangle qubits $q_7$ and $q_{12}$ into a single-one state. Next, two excitation-preserving gates are applied in parallel between qubits $q_7, q_6$, and $q_{12}, q_{11}$. This process can be iterated and the entanglement is propagated towards qubits $q_2$ and $q_{17}$ in parallel. The compiled circuit, including the POVM measurement with neighbouring qubits is shown in Fig.~\ref{fig:w_circuit} of the SM and provided with the accompanying code \cite{github}.

\section{Statistical filter of spurious entanglement}\label{app:statistical_filter}

Given that the maximum likelihood reconstruction of quantum states is generally imperfect, for instance due to finite statistics, it is in principle possible for an experiment to yield non-zero concurrence or negativity despite the underlying state being separable. We thus propose the following method to filter out statistically insignificant, or spurious, entanglement.

We classically simulate the tomographic reconstruction of the following separable but correlated states
\begin{equation}
    \begin{aligned}
        \rho_1 &= \frac{1}{2} \left( \kb{00}{00} + \kb{11}{11}  \right) \\
        \rho_2 &= \frac{1}{2} \left( \kb{000}{000} + \kb{111}{111}  \right)\\
        \rho_3 &= \frac{1}{4} \sum_{i,j=0}^1 \kb{ij}{ij} \otimes \kb{ij}{ij}
    \end{aligned}
\end{equation}
classically, that is, we simulate the noiseless sampling process with the same POVM used in the experiment, with the same number of shots, and we reconstruct the quantum states using the same likelihood maximisation algorithm. This simulation is repeated $10^4$ times for each of the separable input states. For the two-qubit state $\rho_1$, we calculate the concurrence of the resulting density matrices. For $\rho_2$ and $\rho_3$, we calculate negativity according to all possible bipartitions. Notice that the choice of states to generate spurious entanglement statistics is motivated by numerical experiments in which we observed these states to yield the largest values. However, the approach would largely benefit from a more thorough understanding of this phenomenon or exploration of state space. However, this is beyond the scope of this work.

\begin{figure}
\centering
\includegraphics[width=0.92\columnwidth]{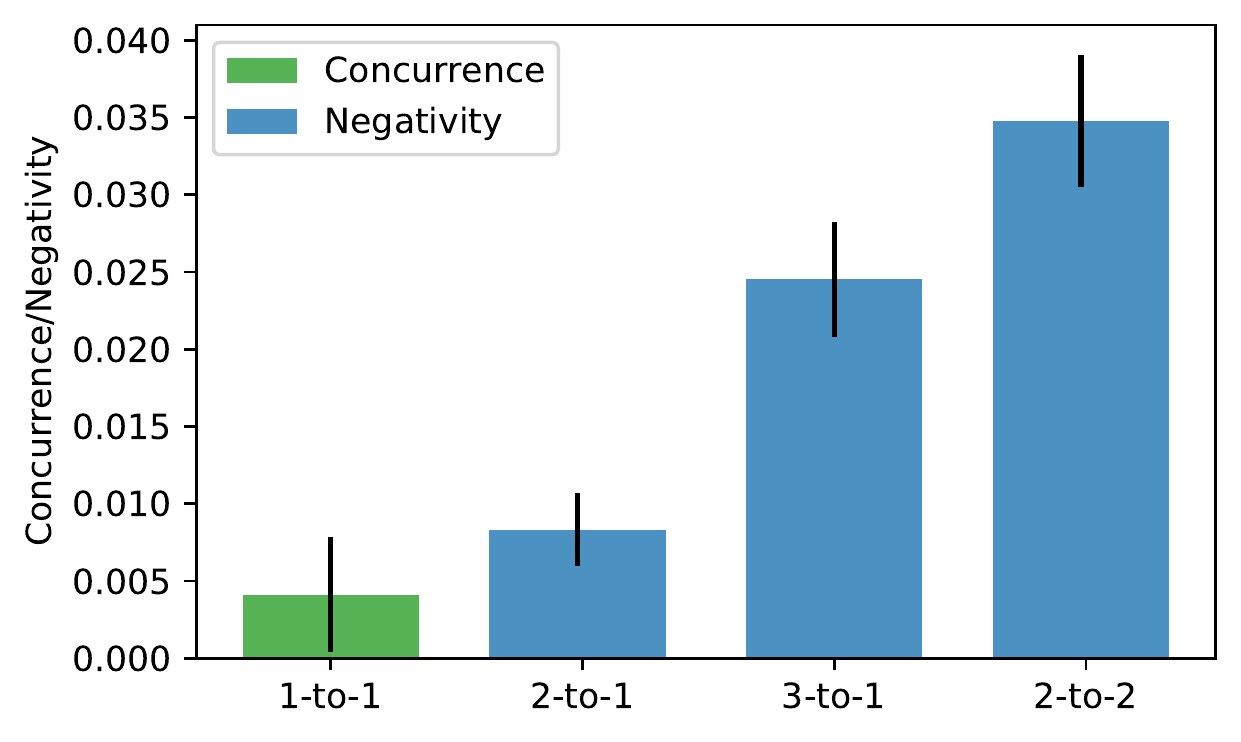}
\caption{Average and standard deviation of the concurrence and negativity of the tomographic reconstruction of states $\rho_1$, $\rho_2$ and $\rho_3$.}
\label{fig:spurious_entanglement}
\end{figure}

In Fig.~\ref{fig:spurious_entanglement}, we show the average and standard deviation of the obtained values. As the simulations show, even fully separable sates can exhibit non-zero negativity due to finite statistics. Based on our data, we can estimate the statistical significance of observed entanglement of up to four-qubit states. If the observed negativity or concurrence is higher than any data point in our simulations, we can give a $p$-value of $10^{-4}$ for the observed entanglement to be legitimate. 

\section{Finding the minimal partitions}\label{app:minimal_partitions}

In this section, we outline the methodology to find the set of minimal partitions $\mathcal{M}_\rho$ consistent with the set of statistically relevant entanglement observations $\mathcal{N}$. Given that the entanglement observations are obtained from bipartitions of reduced $k$-qubit states, $\mathcal{N}$ is composed of sets of the form $\mathcal{Z} = \lbrace Z_1, Z_2 \rbrace$, where the sets $Z_1, Z_2$ fulfil $|Z_1| + |Z_2| = k$. In order for a partition $\mathcal{P}$ to be compatible with a constraint $\mathcal{Z}$, at least two parties $S_i$ and $S_j$, with $S_i \in Z_1$ and $S_j \in Z_2$, must belong to the same set $P \in \mathcal{P}$.

Our strategy is to consider the elements of $\mathcal{N}$ one at a time and keep track of all the minimal partitions compatible up to that point. This means that, when a new element from $\mathcal{N}$ is taken into account, some of the minimal partitions may not be compatible with it and the set must be updated accordingly.

Let the sequence $\mathcal{Z}_1, \ldots, \mathcal{Z}_{|\mathcal{N}|}$ specify some ordering in the set of separability constraints $\mathcal{N}$, and $\mathcal{M}_\rho^{t}$ be the set of minimal partitions compatible with all the constraints $\mathcal{Z}_{t'}$ with $t' \leq t$; initially, when no entanglement is considered, $\mathcal{M}_\rho^{0}$ contains only one partition $\mathcal{P}$ (for which $|P| = 1, \, \forall P \in \mathcal{P}$). The question we now address is how to update $\mathcal{M}_\rho^{t-1}$ to obtain $\mathcal{M}_\rho^{t}$ given $\mathcal{Z}_t$. Notice that if a partition $\mathcal{P} \in \mathcal{M}_\rho^{t-1}$ is compatible with $\mathcal{Z}_t$, it must necessarily be in $\mathcal{M}_\rho^{t}$, as it is compatible with all the so far considered constraints and minimal. If it is not compatible with $\mathcal{Z}_t$, it cannot belong to $\mathcal{M}_\rho^{t}$, but one can easily construct partitions from $\mathcal{P}$ compatible with the constraints and which may be minimal. In particular, let $\mathcal{P} = \lbrace P_1, \ldots, P_{|\mathcal{P}|} \rbrace \in \mathcal{M}_\rho^{t-1}$ and $\mathcal{Z}_t = \lbrace Z_{t,1}, Z_{t,2} \rbrace$, and consider the two sets $\mathcal{A}_j = \lbrace i \vert P_i \cap Z_{t,j} \neq \emptyset \rbrace, j = 1, 2$. Then, any merging of two sets $P_a, P_b \in \mathcal{P}$ with $a \in \mathcal{A}_1$ and $b \in \mathcal{A}_2$ yields a partition $\mathcal{P}'$ compatible with $\mathcal{Z}_{t}$ and all the previous constraints. There are $|\mathcal{A}_1| |\mathcal{A}_2|$ such new partitions $\mathcal{P}'$ that can potentially belong to $\mathcal{M}_\rho^{t}$. However, these partitions are not guaranteed to be minimal, so we must assess whether they are. Given these considerations, we propose the following algorithm:
\begin{itemize}
    \item[1.] At every iteration $t$, define two new sets, $\mathcal{S}$ and $\mathcal{T}$.
    \item[2.] Iterate over all the minimal partitions $\mathcal{P} \in \mathcal{M}_\rho^{t-1}$. For every partition, check if $\mathcal{P}$ is compatible with constraint $\mathcal{Z}_t$.
    \begin{itemize}
        \item[i.] If it is compatible, add $\mathcal{P}$ to $\mathcal{S}$.
        \item[ii.] If it is not, construct the sets $\mathcal{A}_1$ and $\mathcal{A}_2$ and use them to generate the compatible partitions $\mathcal{P}'$, as explained above. Add all these new partitions to $\mathcal{T}$.
    \end{itemize}
    \item[3.] Remove from $\mathcal{T}$ all the partitions for which there exist refinements in $\mathcal{S}$ or $\mathcal{T}$.
\end{itemize}
After these operations, the set of minimal partitions $\mathcal{M}_\rho^{t}$ is given by the union of $\mathcal{S}$ and $\mathcal{T}$. Notice that the motivation for the use of two intermediate sets $\mathcal{S}$ and $\mathcal{T}$ is to avoid the unnecessary verification for the elements in $\mathcal{S}$.

Finally, it is important to stress that the order with which the elements of $\mathcal{N}$ are added can affect the complexity of the algorithm. Currently, the code accompanying this manuscript adds the entanglement observations taking into account the number of qubits $k$ of the reduced state on which it was observed. In particular, it first adds the constraints for $k = 2$, given that adding these always yields $\mathcal{M}_\rho^{t}$ with $|\mathcal{M}_\rho^{t}| = 1$ (since $|\mathcal{A}_1| |\mathcal{A}_2| = 1$). Hence, this criterion potentially minimises the size of the intermediate $\mathcal{M}_\rho^{t}$. However, the specific ordering for observations corresponding to $k > 2$ could in principle be chosen as to make the algorithm more efficient than in its current implementation. Also, notice that in order to find the minimal partitions we can pre-process the set of statistically relevant entanglement observations to remove the redundant ones, that is, those implied by other entanglement observations for smaller $k$, as explained in Sect.~\ref{sec:w_state}.

\begin{figure}
\centering
\includegraphics[width=\columnwidth]{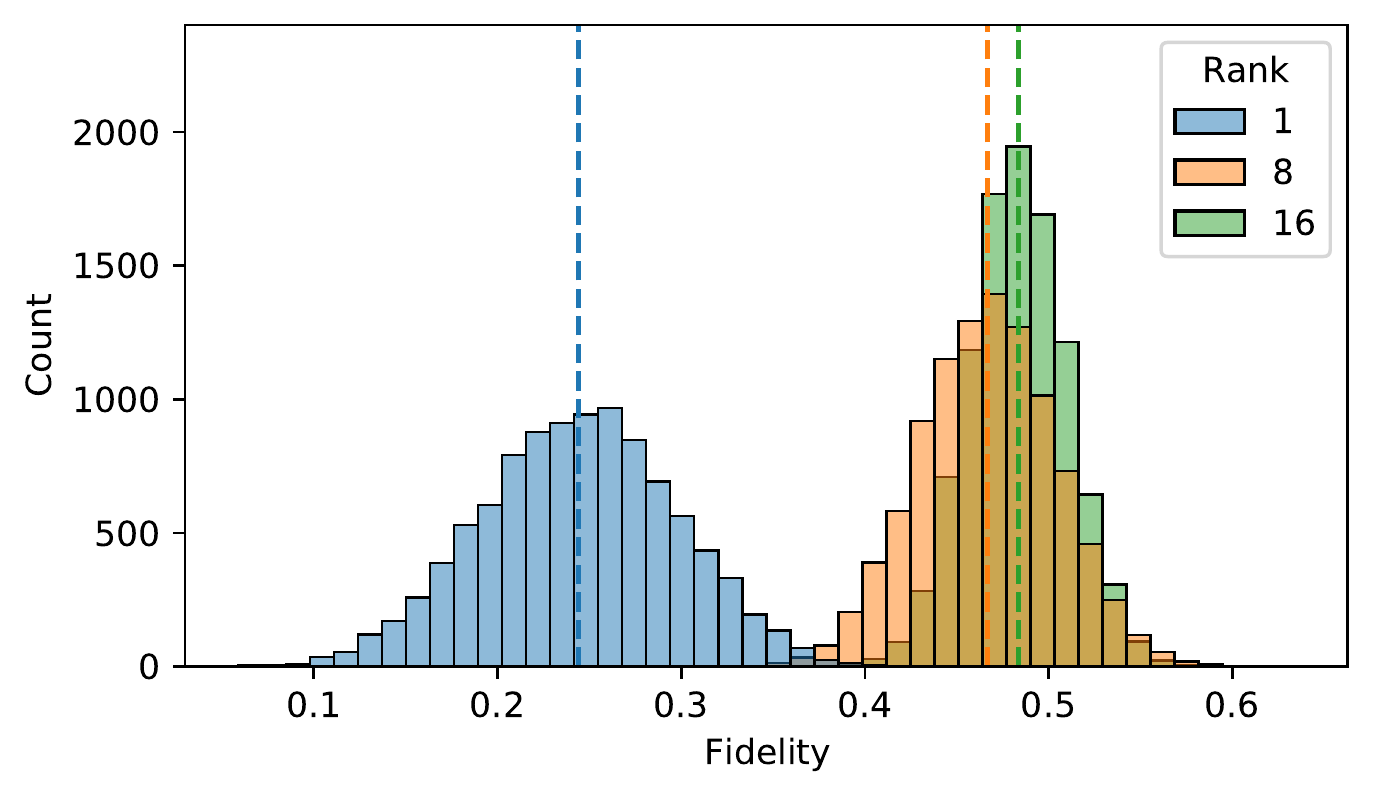}
\caption{Fidelity of randomly sampled states of rank 1, rank 8, and full rank with respect to the Smolin state.}
\label{fig:negativity-rank}
\end{figure}
\section{A note on reconstructed fidelity}\label{app:fidelity}

Our results for the Smolin state reveal the interesting fact that the experimentally reconstructed state presents the level-I separability structure expected from the theory despite the low fidelity. While a fidelity of 0.64 clearly signals that the two states cannot be considered similar, it is interesting to see that such fidelity with respect to the Smolin state is not typical in state space.

In Fig.~\ref{fig:negativity-rank} we show the distribution of the fidelity with respect of the Smolin state of $10^{4}$ Ginibre random states of rank 1, 8 and 16. We can see that the values are distributed approximately according to the normal distribution. Interestingly, the standard deviation of the fidelity decreases as the rank of the random state is increased to full rank. At the same time, the largest average fidelity is obtained for the full rank states. From these distributions, it is easy to see that a value of 0.64 is highly unlikely, which is consistent with the observation that the experimentally reconstructed state, despite the numerous sources of noise, retains some of the expected properties.

\bibliography{bibliography.bib}

\begin{thebibliography}{32}%
\makeatletter
\providecommand \@ifxundefined [1]{%
 \@ifx{#1\undefined}
}%
\providecommand \@ifnum [1]{%
 \ifnum #1\expandafter \@firstoftwo
 \else \expandafter \@secondoftwo
 \fi
}%
\providecommand \@ifx [1]{%
 \ifx #1\expandafter \@firstoftwo
 \else \expandafter \@secondoftwo
 \fi
}%
\providecommand \natexlab [1]{#1}%
\providecommand \enquote  [1]{``#1''}%
\providecommand \bibnamefont  [1]{#1}%
\providecommand \bibfnamefont [1]{#1}%
\providecommand \citenamefont [1]{#1}%
\providecommand \href@noop [0]{\@secondoftwo}%
\providecommand \href [0]{\begingroup \@sanitize@url \@href}%
\providecommand \@href[1]{\@@startlink{#1}\@@href}%
\providecommand \@@href[1]{\endgroup#1\@@endlink}%
\providecommand \@sanitize@url [0]{\catcode `\\12\catcode `\$12\catcode
  `\&12\catcode `\#12\catcode `\^12\catcode `\_12\catcode `\%12\relax}%
\providecommand \@@startlink[1]{}%
\providecommand \@@endlink[0]{}%
\providecommand \url  [0]{\begingroup\@sanitize@url \@url }%
\providecommand \@url [1]{\endgroup\@href {#1}{\urlprefix }}%
\providecommand \urlprefix  [0]{URL }%
\providecommand \Eprint [0]{\href }%
\providecommand \doibase [0]{http://dx.doi.org/}%
\providecommand \selectlanguage [0]{\@gobble}%
\providecommand \bibinfo  [0]{\@secondoftwo}%
\providecommand \bibfield  [0]{\@secondoftwo}%
\providecommand \translation [1]{[#1]}%
\providecommand \BibitemOpen [0]{}%
\providecommand \bibitemStop [0]{}%
\providecommand \bibitemNoStop [0]{.\EOS\space}%
\providecommand \EOS [0]{\spacefactor3000\relax}%
\providecommand \BibitemShut  [1]{\csname bibitem#1\endcsname}%
\let\auto@bib@innerbib\@empty
\bibitem [{\citenamefont {Amico}\ \emph {et~al.}(2008)\citenamefont {Amico},
  \citenamefont {Fazio}, \citenamefont {Osterloh},\ and\ \citenamefont
  {Vedral}}]{Amico2008}%
  \BibitemOpen
  \bibfield  {author} {\bibinfo {author} {\bibfnamefont {Luigi}\ \bibnamefont
  {Amico}}, \bibinfo {author} {\bibfnamefont {Rosario}\ \bibnamefont {Fazio}},
  \bibinfo {author} {\bibfnamefont {Andreas}\ \bibnamefont {Osterloh}}, \ and\
  \bibinfo {author} {\bibfnamefont {Vlatko}\ \bibnamefont {Vedral}},\
  }\bibfield  {title} {\enquote {\bibinfo {title} {Entanglement in many-body
  systems},}\ }\href {\doibase 10.1103/RevModPhys.80.517} {\bibfield  {journal}
  {\bibinfo  {journal} {Rev. Mod. Phys.}\ }\textbf {\bibinfo {volume} {80}},\
  \bibinfo {pages} {517--576} (\bibinfo {year} {2008})}\BibitemShut {NoStop}%
\bibitem [{\citenamefont {Horodecki}\ \emph {et~al.}(2009)\citenamefont
  {Horodecki}, \citenamefont {Horodecki}, \citenamefont {Horodecki},\ and\
  \citenamefont {Horodecki}}]{HorodeckiRyszard2009Qe}%
  \BibitemOpen
  \bibfield  {author} {\bibinfo {author} {\bibfnamefont {Ryszard}\ \bibnamefont
  {Horodecki}}, \bibinfo {author} {\bibfnamefont {Pawe\l{}}\ \bibnamefont
  {Horodecki}}, \bibinfo {author} {\bibfnamefont {Micha\l{}}\ \bibnamefont
  {Horodecki}}, \ and\ \bibinfo {author} {\bibfnamefont {Karol}\ \bibnamefont
  {Horodecki}},\ }\bibfield  {title} {\enquote {\bibinfo {title} {Quantum
  entanglement},}\ }\href {\doibase 10.1103/RevModPhys.81.865} {\bibfield
  {journal} {\bibinfo  {journal} {Rev. Mod. Phys.}\ }\textbf {\bibinfo {volume}
  {81}},\ \bibinfo {pages} {865--942} (\bibinfo {year} {2009})}\BibitemShut
  {NoStop}%
\bibitem [{\citenamefont {G\"{u}hne}\ and\ \citenamefont
  {T{\'{o}}th}(2009)}]{Guhne2009}%
  \BibitemOpen
  \bibfield  {author} {\bibinfo {author} {\bibfnamefont {Otfried}\ \bibnamefont
  {G\"{u}hne}}\ and\ \bibinfo {author} {\bibfnamefont {G{\'{e}}za}\
  \bibnamefont {T{\'{o}}th}},\ }\bibfield  {title} {\enquote {\bibinfo {title}
  {Entanglement detection},}\ }\href {\doibase 10.1016/j.physrep.2009.02.004}
  {\bibfield  {journal} {\bibinfo  {journal} {Physics Reports}\ }\textbf
  {\bibinfo {volume} {474}},\ \bibinfo {pages} {1--75} (\bibinfo {year}
  {2009})}\BibitemShut {NoStop}%
\bibitem [{\citenamefont {Modi}\ \emph {et~al.}(2012)\citenamefont {Modi},
  \citenamefont {Brodutch}, \citenamefont {Cable}, \citenamefont {Paterek},\
  and\ \citenamefont {Vedral}}]{Modi2012}%
  \BibitemOpen
  \bibfield  {author} {\bibinfo {author} {\bibfnamefont {Kavan}\ \bibnamefont
  {Modi}}, \bibinfo {author} {\bibfnamefont {Aharon}\ \bibnamefont {Brodutch}},
  \bibinfo {author} {\bibfnamefont {Hugo}\ \bibnamefont {Cable}}, \bibinfo
  {author} {\bibfnamefont {Tomasz}\ \bibnamefont {Paterek}}, \ and\ \bibinfo
  {author} {\bibfnamefont {Vlatko}\ \bibnamefont {Vedral}},\ }\bibfield
  {title} {\enquote {\bibinfo {title} {The classical-quantum boundary for
  correlations: Discord and related measures},}\ }\href {\doibase
  10.1103/RevModPhys.84.1655} {\bibfield  {journal} {\bibinfo  {journal} {Rev.
  Mod. Phys.}\ }\textbf {\bibinfo {volume} {84}},\ \bibinfo {pages}
  {1655--1707} (\bibinfo {year} {2012})}\BibitemShut {NoStop}%
\bibitem [{\citenamefont {Adesso}\ \emph {et~al.}(2016)\citenamefont {Adesso},
  \citenamefont {Bromley},\ and\ \citenamefont {Cianciaruso}}]{Adesso2016}%
  \BibitemOpen
  \bibfield  {author} {\bibinfo {author} {\bibfnamefont {Gerardo}\ \bibnamefont
  {Adesso}}, \bibinfo {author} {\bibfnamefont {Thomas~R}\ \bibnamefont
  {Bromley}}, \ and\ \bibinfo {author} {\bibfnamefont {Marco}\ \bibnamefont
  {Cianciaruso}},\ }\bibfield  {title} {\enquote {\bibinfo {title} {Measures
  and applications of quantum correlations},}\ }\href {\doibase
  10.1088/1751-8113/49/47/473001} {\bibfield  {journal} {\bibinfo  {journal}
  {Journal of Physics A: Mathematical and Theoretical}\ }\textbf {\bibinfo
  {volume} {49}},\ \bibinfo {pages} {473001} (\bibinfo {year}
  {2016})}\BibitemShut {NoStop}%
\bibitem [{\citenamefont {Hebenstreit}\ \emph {et~al.}(2016)\citenamefont
  {Hebenstreit}, \citenamefont {Spee},\ and\ \citenamefont
  {Kraus}}]{Hebenstreit2016}%
  \BibitemOpen
  \bibfield  {author} {\bibinfo {author} {\bibfnamefont {M.}~\bibnamefont
  {Hebenstreit}}, \bibinfo {author} {\bibfnamefont {C.}~\bibnamefont {Spee}}, \
  and\ \bibinfo {author} {\bibfnamefont {B.}~\bibnamefont {Kraus}},\ }\bibfield
   {title} {\enquote {\bibinfo {title} {Maximally entangled set of tripartite
  qutrit states and pure state separable transformations which are not possible
  via local operations and classical communication},}\ }\href {\doibase
  10.1103/PhysRevA.93.012339} {\bibfield  {journal} {\bibinfo  {journal} {Phys.
  Rev. A}\ }\textbf {\bibinfo {volume} {93}},\ \bibinfo {pages} {012339}
  (\bibinfo {year} {2016})}\BibitemShut {NoStop}%
\bibitem [{\citenamefont {Sauerwein}\ \emph {et~al.}(2018)\citenamefont
  {Sauerwein}, \citenamefont {Wallach}, \citenamefont {Gour},\ and\
  \citenamefont {Kraus}}]{Sauerwein2018}%
  \BibitemOpen
  \bibfield  {author} {\bibinfo {author} {\bibfnamefont {David}\ \bibnamefont
  {Sauerwein}}, \bibinfo {author} {\bibfnamefont {Nolan~R.}\ \bibnamefont
  {Wallach}}, \bibinfo {author} {\bibfnamefont {Gilad}\ \bibnamefont {Gour}}, \
  and\ \bibinfo {author} {\bibfnamefont {Barbara}\ \bibnamefont {Kraus}},\
  }\bibfield  {title} {\enquote {\bibinfo {title} {Transformations among pure
  multipartite entangled states via local operations are almost never
  possible},}\ }\href {\doibase 10.1103/PhysRevX.8.031020} {\bibfield
  {journal} {\bibinfo  {journal} {Phys. Rev. X}\ }\textbf {\bibinfo {volume}
  {8}},\ \bibinfo {pages} {031020} (\bibinfo {year} {2018})}\BibitemShut
  {NoStop}%
\bibitem [{\citenamefont {Szalay}\ and\ \citenamefont
  {K\"ok\'enyesi}(2012)}]{Szalay2012}%
  \BibitemOpen
  \bibfield  {author} {\bibinfo {author} {\bibfnamefont {Szil\'ard}\
  \bibnamefont {Szalay}}\ and\ \bibinfo {author} {\bibfnamefont {Zolt\'an}\
  \bibnamefont {K\"ok\'enyesi}},\ }\bibfield  {title} {\enquote {\bibinfo
  {title} {Partial separability revisited: Necessary and sufficient
  criteria},}\ }\href {\doibase 10.1103/PhysRevA.86.032341} {\bibfield
  {journal} {\bibinfo  {journal} {Phys. Rev. A}\ }\textbf {\bibinfo {volume}
  {86}},\ \bibinfo {pages} {032341} (\bibinfo {year} {2012})}\BibitemShut
  {NoStop}%
\bibitem [{\citenamefont {Szalay}(2015)}]{szalay2015multipartite}%
  \BibitemOpen
  \bibfield  {author} {\bibinfo {author} {\bibfnamefont {Szil\'ard}\
  \bibnamefont {Szalay}},\ }\bibfield  {title} {\enquote {\bibinfo {title}
  {Multipartite entanglement measures},}\ }\href {\doibase
  10.1103/PhysRevA.92.042329} {\bibfield  {journal} {\bibinfo  {journal} {Phys.
  Rev. A}\ }\textbf {\bibinfo {volume} {92}},\ \bibinfo {pages} {042329}
  (\bibinfo {year} {2015})}\BibitemShut {NoStop}%
\bibitem [{\citenamefont {Szalay}(2018)}]{Szalay2018}%
  \BibitemOpen
  \bibfield  {author} {\bibinfo {author} {\bibfnamefont {Szil{\'{a}}rd}\
  \bibnamefont {Szalay}},\ }\bibfield  {title} {\enquote {\bibinfo {title} {The
  classification of multipartite quantum correlation},}\ }\href {\doibase
  10.1088/1751-8121/aae971} {\bibfield  {journal} {\bibinfo  {journal} {Journal
  of Physics A: Mathematical and Theoretical}\ }\textbf {\bibinfo {volume}
  {51}},\ \bibinfo {pages} {485302} (\bibinfo {year} {2018})}\BibitemShut
  {NoStop}%
\bibitem [{\citenamefont {Szalay}(2019)}]{SzalaySzilard2019koea}%
  \BibitemOpen
  \bibfield  {author} {\bibinfo {author} {\bibfnamefont {Szil\'ard}\
  \bibnamefont {Szalay}},\ }\bibfield  {title} {\enquote {\bibinfo {title}
  {k-stretchability of entanglement, and the duality of k-separability and
  k-producibility},}\ }\href {\doibase 10.22331/q-2019-12-02-204} {\bibfield
  {journal} {\bibinfo  {journal} {Quantum}\ }\textbf {\bibinfo {volume} {3}},\
  \bibinfo {pages} {204} (\bibinfo {year} {2019})}\BibitemShut {NoStop}%
\bibitem [{\citenamefont {DiVincenzo}\ \emph {et~al.}(2003)\citenamefont
  {DiVincenzo}, \citenamefont {Mor}, \citenamefont {Shor}, \citenamefont
  {Smolin},\ and\ \citenamefont {Terhal}}]{DIVINCENZODavidP2003Upbu}%
  \BibitemOpen
  \bibfield  {author} {\bibinfo {author} {\bibfnamefont {David~P.}\
  \bibnamefont {DiVincenzo}}, \bibinfo {author} {\bibfnamefont {Tal}\
  \bibnamefont {Mor}}, \bibinfo {author} {\bibfnamefont {Peter~W.}\
  \bibnamefont {Shor}}, \bibinfo {author} {\bibfnamefont {John~A.}\
  \bibnamefont {Smolin}}, \ and\ \bibinfo {author} {\bibfnamefont {Barbara~M.}\
  \bibnamefont {Terhal}},\ }\bibfield  {title} {\enquote {\bibinfo {title}
  {Unextendible product bases, uncompletable product bases and bound
  entanglement},}\ }\href {\doibase 10.1007/s00220-003-0877-6} {\bibfield
  {journal} {\bibinfo  {journal} {Communications in Mathematical Physics}\
  }\textbf {\bibinfo {volume} {238}},\ \bibinfo {pages} {379--410} (\bibinfo
  {year} {2003})}\BibitemShut {NoStop}%
\bibitem [{\citenamefont {Jungnitsch}\ \emph {et~al.}(2011)\citenamefont
  {Jungnitsch}, \citenamefont {Moroder},\ and\ \citenamefont
  {G\"uhne}}]{TamingMultiparticleEntanglement}%
  \BibitemOpen
  \bibfield  {author} {\bibinfo {author} {\bibfnamefont {Bastian}\ \bibnamefont
  {Jungnitsch}}, \bibinfo {author} {\bibfnamefont {Tobias}\ \bibnamefont
  {Moroder}}, \ and\ \bibinfo {author} {\bibfnamefont {Otfried}\ \bibnamefont
  {G\"uhne}},\ }\bibfield  {title} {\enquote {\bibinfo {title} {Taming
  multiparticle entanglement},}\ }\href {\doibase
  10.1103/PhysRevLett.106.190502} {\bibfield  {journal} {\bibinfo  {journal}
  {Phys. Rev. Lett.}\ }\textbf {\bibinfo {volume} {106}},\ \bibinfo {pages}
  {190502} (\bibinfo {year} {2011})}\BibitemShut {NoStop}%
\bibitem [{\citenamefont {Friis}\ \emph
  {et~al.}(2018{\natexlab{a}})\citenamefont {Friis}, \citenamefont {Marty},
  \citenamefont {Maier}, \citenamefont {Hempel}, \citenamefont {Holz\"apfel},
  \citenamefont {Jurcevic}, \citenamefont {Plenio}, \citenamefont {Huber},
  \citenamefont {Roos}, \citenamefont {Blatt},\ and\ \citenamefont
  {Lanyon}}]{Friis2018}%
  \BibitemOpen
  \bibfield  {author} {\bibinfo {author} {\bibfnamefont {Nicolai}\ \bibnamefont
  {Friis}}, \bibinfo {author} {\bibfnamefont {Oliver}\ \bibnamefont {Marty}},
  \bibinfo {author} {\bibfnamefont {Christine}\ \bibnamefont {Maier}}, \bibinfo
  {author} {\bibfnamefont {Cornelius}\ \bibnamefont {Hempel}}, \bibinfo
  {author} {\bibfnamefont {Milan}\ \bibnamefont {Holz\"apfel}}, \bibinfo
  {author} {\bibfnamefont {Petar}\ \bibnamefont {Jurcevic}}, \bibinfo {author}
  {\bibfnamefont {Martin~B.}\ \bibnamefont {Plenio}}, \bibinfo {author}
  {\bibfnamefont {Marcus}\ \bibnamefont {Huber}}, \bibinfo {author}
  {\bibfnamefont {Christian}\ \bibnamefont {Roos}}, \bibinfo {author}
  {\bibfnamefont {Rainer}\ \bibnamefont {Blatt}}, \ and\ \bibinfo {author}
  {\bibfnamefont {Ben}\ \bibnamefont {Lanyon}},\ }\bibfield  {title} {\enquote
  {\bibinfo {title} {Observation of entangled states of a fully controlled
  20-qubit system},}\ }\href {\doibase 10.1103/PhysRevX.8.021012} {\bibfield
  {journal} {\bibinfo  {journal} {Phys. Rev. X}\ }\textbf {\bibinfo {volume}
  {8}},\ \bibinfo {pages} {021012} (\bibinfo {year}
  {2018}{\natexlab{a}})}\BibitemShut {NoStop}%
\bibitem [{\citenamefont {Lu}\ \emph {et~al.}(2018)\citenamefont {Lu},
  \citenamefont {Zhao}, \citenamefont {Li}, \citenamefont {Yin}, \citenamefont
  {Yuan}, \citenamefont {Hung}, \citenamefont {Chen}, \citenamefont {Li},
  \citenamefont {Liu}, \citenamefont {Peng}, \citenamefont {Liang},
  \citenamefont {Ma}, \citenamefont {Chen},\ and\ \citenamefont
  {Pan}}]{Lu2018}%
  \BibitemOpen
  \bibfield  {author} {\bibinfo {author} {\bibfnamefont {He}~\bibnamefont
  {Lu}}, \bibinfo {author} {\bibfnamefont {Qi}~\bibnamefont {Zhao}}, \bibinfo
  {author} {\bibfnamefont {Zheng-Da}\ \bibnamefont {Li}}, \bibinfo {author}
  {\bibfnamefont {Xu-Fei}\ \bibnamefont {Yin}}, \bibinfo {author}
  {\bibfnamefont {Xiao}\ \bibnamefont {Yuan}}, \bibinfo {author} {\bibfnamefont
  {Jui-Chen}\ \bibnamefont {Hung}}, \bibinfo {author} {\bibfnamefont {Luo-Kan}\
  \bibnamefont {Chen}}, \bibinfo {author} {\bibfnamefont {Li}~\bibnamefont
  {Li}}, \bibinfo {author} {\bibfnamefont {Nai-Le}\ \bibnamefont {Liu}},
  \bibinfo {author} {\bibfnamefont {Cheng-Zhi}\ \bibnamefont {Peng}}, \bibinfo
  {author} {\bibfnamefont {Yeong-Cherng}\ \bibnamefont {Liang}}, \bibinfo
  {author} {\bibfnamefont {Xiongfeng}\ \bibnamefont {Ma}}, \bibinfo {author}
  {\bibfnamefont {Yu-Ao}\ \bibnamefont {Chen}}, \ and\ \bibinfo {author}
  {\bibfnamefont {Jian-Wei}\ \bibnamefont {Pan}},\ }\bibfield  {title}
  {\enquote {\bibinfo {title} {Entanglement structure: Entanglement
  partitioning in multipartite systems and its experimental detection using
  optimizable witnesses},}\ }\href {\doibase 10.1103/PhysRevX.8.021072}
  {\bibfield  {journal} {\bibinfo  {journal} {Phys. Rev. X}\ }\textbf {\bibinfo
  {volume} {8}},\ \bibinfo {pages} {021072} (\bibinfo {year}
  {2018})}\BibitemShut {NoStop}%
\bibitem [{\citenamefont {Saggio}\ \emph {et~al.}(2019)\citenamefont {Saggio},
  \citenamefont {Dimi{\'{c}}}, \citenamefont {Greganti}, \citenamefont
  {Rozema}, \citenamefont {Walther},\ and\ \citenamefont
  {Daki{\'{c}}}}]{Saggio2019}%
  \BibitemOpen
  \bibfield  {author} {\bibinfo {author} {\bibfnamefont {Valeria}\ \bibnamefont
  {Saggio}}, \bibinfo {author} {\bibfnamefont {Aleksandra}\ \bibnamefont
  {Dimi{\'{c}}}}, \bibinfo {author} {\bibfnamefont {Chiara}\ \bibnamefont
  {Greganti}}, \bibinfo {author} {\bibfnamefont {Lee~A.}\ \bibnamefont
  {Rozema}}, \bibinfo {author} {\bibfnamefont {Philip}\ \bibnamefont
  {Walther}}, \ and\ \bibinfo {author} {\bibfnamefont {Borivoje}\ \bibnamefont
  {Daki{\'{c}}}},\ }\bibfield  {title} {\enquote {\bibinfo {title}
  {Experimental few-copy multipartite entanglement detection},}\ }\href
  {\doibase 10.1038/s41567-019-0550-4} {\bibfield  {journal} {\bibinfo
  {journal} {Nature Physics}\ }\textbf {\bibinfo {volume} {15}},\ \bibinfo
  {pages} {935--940} (\bibinfo {year} {2019})}\BibitemShut {NoStop}%
\bibitem [{\citenamefont {Friis}\ \emph
  {et~al.}(2018{\natexlab{b}})\citenamefont {Friis}, \citenamefont
  {Vitagliano}, \citenamefont {Malik},\ and\ \citenamefont
  {Huber}}]{Friis2018review}%
  \BibitemOpen
  \bibfield  {author} {\bibinfo {author} {\bibfnamefont {Nicolai}\ \bibnamefont
  {Friis}}, \bibinfo {author} {\bibfnamefont {Giuseppe}\ \bibnamefont
  {Vitagliano}}, \bibinfo {author} {\bibfnamefont {Mehul}\ \bibnamefont
  {Malik}}, \ and\ \bibinfo {author} {\bibfnamefont {Marcus}\ \bibnamefont
  {Huber}},\ }\bibfield  {title} {\enquote {\bibinfo {title} {Entanglement
  certification from theory to experiment},}\ }\href {\doibase
  10.1038/s42254-018-0003-5} {\bibfield  {journal} {\bibinfo  {journal} {Nature
  Reviews Physics}\ }\textbf {\bibinfo {volume} {1}},\ \bibinfo {pages}
  {72--87} (\bibinfo {year} {2018}{\natexlab{b}})}\BibitemShut {NoStop}%
\bibitem [{\citenamefont {Smolin}(2001)}]{Smolin2001}%
  \BibitemOpen
  \bibfield  {author} {\bibinfo {author} {\bibfnamefont {John~A.}\ \bibnamefont
  {Smolin}},\ }\bibfield  {title} {\enquote {\bibinfo {title} {Four-party
  unlockable bound entangled state},}\ }\href {\doibase
  10.1103/PhysRevA.63.032306} {\bibfield  {journal} {\bibinfo  {journal} {Phys.
  Rev. A}\ }\textbf {\bibinfo {volume} {63}},\ \bibinfo {pages} {032306}
  (\bibinfo {year} {2001})}\BibitemShut {NoStop}%
\bibitem [{\citenamefont {Cotler}\ and\ \citenamefont
  {Wilczek}(2020)}]{CotelrPRL2020}%
  \BibitemOpen
  \bibfield  {author} {\bibinfo {author} {\bibfnamefont {Jordan}\ \bibnamefont
  {Cotler}}\ and\ \bibinfo {author} {\bibfnamefont {Frank}\ \bibnamefont
  {Wilczek}},\ }\bibfield  {title} {\enquote {\bibinfo {title} {Quantum
  overlapping tomography},}\ }\href {\doibase 10.1103/PhysRevLett.124.100401}
  {\bibfield  {journal} {\bibinfo  {journal} {Phys. Rev. Lett.}\ }\textbf
  {\bibinfo {volume} {124}},\ \bibinfo {pages} {100401} (\bibinfo {year}
  {2020})}\BibitemShut {NoStop}%
\bibitem [{\citenamefont {Garc{\'{i}}a-P{\'{e}}rez}\ \emph
  {et~al.}(2020)\citenamefont {Garc{\'{i}}a-P{\'{e}}rez}, \citenamefont
  {Rossi}, \citenamefont {Sokolov}, \citenamefont {Borrelli},\ and\
  \citenamefont {Maniscalco}}]{Garcia-Perez2020}%
  \BibitemOpen
  \bibfield  {author} {\bibinfo {author} {\bibfnamefont {Guillermo}\
  \bibnamefont {Garc{\'{i}}a-P{\'{e}}rez}}, \bibinfo {author} {\bibfnamefont
  {Matteo A.~C.}\ \bibnamefont {Rossi}}, \bibinfo {author} {\bibfnamefont
  {Boris}\ \bibnamefont {Sokolov}}, \bibinfo {author} {\bibfnamefont
  {Elsi-Mari}\ \bibnamefont {Borrelli}}, \ and\ \bibinfo {author}
  {\bibfnamefont {Sabrina}\ \bibnamefont {Maniscalco}},\ }\bibfield  {title}
  {\enquote {\bibinfo {title} {{Pairwise tomography networks for many-body
  quantum systems}},}\ }\href {\doibase 10.1103/PhysRevResearch.2.023393}
  {\bibfield  {journal} {\bibinfo  {journal} {Phys. Rev. Res.}\ }\textbf
  {\bibinfo {volume} {2}},\ \bibinfo {pages} {023393} (\bibinfo {year}
  {2020})},\ \Eprint {http://arxiv.org/abs/1909.12814} {arXiv:1909.12814}
  \BibitemShut {NoStop}%
\bibitem [{\citenamefont {Bonet-Monroig}\ \emph {et~al.}(2020)\citenamefont
  {Bonet-Monroig}, \citenamefont {Babbush},\ and\ \citenamefont
  {O'Brien}}]{bonet-monroig2020nearly}%
  \BibitemOpen
  \bibfield  {author} {\bibinfo {author} {\bibfnamefont {Xavier}\ \bibnamefont
  {Bonet-Monroig}}, \bibinfo {author} {\bibfnamefont {Ryan}\ \bibnamefont
  {Babbush}}, \ and\ \bibinfo {author} {\bibfnamefont {Thomas~E.}\ \bibnamefont
  {O'Brien}},\ }\bibfield  {title} {\enquote {\bibinfo {title} {Nearly optimal
  measurement scheduling for partial tomography of quantum states},}\ }\href
  {\doibase 10.1103/PhysRevX.10.031064} {\bibfield  {journal} {\bibinfo
  {journal} {Phys. Rev. X}\ }\textbf {\bibinfo {volume} {10}},\ \bibinfo
  {pages} {031064} (\bibinfo {year} {2020})}\BibitemShut {NoStop}%
\bibitem [{\citenamefont {Jiang}\ \emph {et~al.}(2020)\citenamefont {Jiang},
  \citenamefont {Kalev}, \citenamefont {Mruczkiewicz},\ and\ \citenamefont
  {Neven}}]{Jiang2020optimalfermionto}%
  \BibitemOpen
  \bibfield  {author} {\bibinfo {author} {\bibfnamefont {Zhang}\ \bibnamefont
  {Jiang}}, \bibinfo {author} {\bibfnamefont {Amir}\ \bibnamefont {Kalev}},
  \bibinfo {author} {\bibfnamefont {Wojciech}\ \bibnamefont {Mruczkiewicz}}, \
  and\ \bibinfo {author} {\bibfnamefont {Hartmut}\ \bibnamefont {Neven}},\
  }\bibfield  {title} {\enquote {\bibinfo {title} {Optimal fermion-to-qubit
  mapping via ternary trees with applications to reduced quantum states
  learning},}\ }\href {\doibase 10.22331/q-2020-06-04-276} {\bibfield
  {journal} {\bibinfo  {journal} {{Quantum}}\ }\textbf {\bibinfo {volume}
  {4}},\ \bibinfo {pages} {276} (\bibinfo {year} {2020})}\BibitemShut {NoStop}%
\bibitem [{\citenamefont {Hradil}(1997)}]{Hradil1997quantumstate}%
  \BibitemOpen
  \bibfield  {author} {\bibinfo {author} {\bibfnamefont {Z.}~\bibnamefont
  {Hradil}},\ }\bibfield  {title} {\enquote {\bibinfo {title} {Quantum-state
  estimation},}\ }\href {\doibase 10.1103/PhysRevA.55.R1561} {\bibfield
  {journal} {\bibinfo  {journal} {Phys. Rev. A}\ }\textbf {\bibinfo {volume}
  {55}},\ \bibinfo {pages} {R1561--R1564} (\bibinfo {year} {1997})}\BibitemShut
  {NoStop}%
\bibitem [{\citenamefont {Banaszek}\ \emph {et~al.}(1999)\citenamefont
  {Banaszek}, \citenamefont {D'Ariano}, \citenamefont {Paris},\ and\
  \citenamefont {Sacchi}}]{banaszek1999maximum}%
  \BibitemOpen
  \bibfield  {author} {\bibinfo {author} {\bibfnamefont {K.}~\bibnamefont
  {Banaszek}}, \bibinfo {author} {\bibfnamefont {G.~M.}\ \bibnamefont
  {D'Ariano}}, \bibinfo {author} {\bibfnamefont {M.~G.~A.}\ \bibnamefont
  {Paris}}, \ and\ \bibinfo {author} {\bibfnamefont {M.~F.}\ \bibnamefont
  {Sacchi}},\ }\bibfield  {title} {\enquote {\bibinfo {title}
  {Maximum-likelihood estimation of the density matrix},}\ }\href {\doibase
  10.1103/PhysRevA.61.010304} {\bibfield  {journal} {\bibinfo  {journal} {Phys.
  Rev. A}\ }\textbf {\bibinfo {volume} {61}},\ \bibinfo {pages} {010304}
  (\bibinfo {year} {1999})}\BibitemShut {NoStop}%
\bibitem [{\citenamefont {Smolin}\ \emph {et~al.}(2012)\citenamefont {Smolin},
  \citenamefont {Gambetta},\ and\ \citenamefont {Smith}}]{smolin2012efficient}%
  \BibitemOpen
  \bibfield  {author} {\bibinfo {author} {\bibfnamefont {John~A.}\ \bibnamefont
  {Smolin}}, \bibinfo {author} {\bibfnamefont {Jay~M.}\ \bibnamefont
  {Gambetta}}, \ and\ \bibinfo {author} {\bibfnamefont {Graeme}\ \bibnamefont
  {Smith}},\ }\bibfield  {title} {\enquote {\bibinfo {title} {Efficient method
  for computing the maximum-likelihood quantum state from measurements with
  additive gaussian noise},}\ }\href {\doibase 10.1103/PhysRevLett.108.070502}
  {\bibfield  {journal} {\bibinfo  {journal} {Phys. Rev. Lett.}\ }\textbf
  {\bibinfo {volume} {108}},\ \bibinfo {pages} {070502} (\bibinfo {year}
  {2012})}\BibitemShut {NoStop}%
\bibitem [{\citenamefont {\ifmmode \check{R}\else
  \v{R}\fi{}eh\'a\ifmmode~\check{c}\else \v{c}\fi{}ek}\ \emph
  {et~al.}(2007)\citenamefont {\ifmmode \check{R}\else
  \v{R}\fi{}eh\'a\ifmmode~\check{c}\else \v{c}\fi{}ek}, \citenamefont {Hradil},
  \citenamefont {Knill},\ and\ \citenamefont {Lvovsky}}]{Rehacek2007diluted}%
  \BibitemOpen
  \bibfield  {author} {\bibinfo {author} {\bibfnamefont {Jaroslav}\
  \bibnamefont {\ifmmode \check{R}\else \v{R}\fi{}eh\'a\ifmmode~\check{c}\else
  \v{c}\fi{}ek}}, \bibinfo {author} {\bibfnamefont {Zden\v{e}k}\ \bibnamefont
  {Hradil}}, \bibinfo {author} {\bibfnamefont {E.}~\bibnamefont {Knill}}, \
  and\ \bibinfo {author} {\bibfnamefont {A.~I.}\ \bibnamefont {Lvovsky}},\
  }\bibfield  {title} {\enquote {\bibinfo {title} {Diluted maximum-likelihood
  algorithm for quantum tomography},}\ }\href {\doibase
  10.1103/PhysRevA.75.042108} {\bibfield  {journal} {\bibinfo  {journal} {Phys.
  Rev. A}\ }\textbf {\bibinfo {volume} {75}},\ \bibinfo {pages} {042108}
  (\bibinfo {year} {2007})}\BibitemShut {NoStop}%
\bibitem [{\citenamefont {Wootters}(1998)}]{WoottersConcurrence}%
  \BibitemOpen
  \bibfield  {author} {\bibinfo {author} {\bibfnamefont {William~K.}\
  \bibnamefont {Wootters}},\ }\bibfield  {title} {\enquote {\bibinfo {title}
  {Entanglement of formation of an arbitrary state of two qubits},}\ }\href
  {\doibase 10.1103/PhysRevLett.80.2245} {\bibfield  {journal} {\bibinfo
  {journal} {Phys. Rev. Lett.}\ }\textbf {\bibinfo {volume} {80}},\ \bibinfo
  {pages} {2245--2248} (\bibinfo {year} {1998})}\BibitemShut {NoStop}%
\bibitem [{\citenamefont {Vidal}\ and\ \citenamefont
  {Werner}(2002)}]{VidalWernerNegativity}%
  \BibitemOpen
  \bibfield  {author} {\bibinfo {author} {\bibfnamefont {G.}~\bibnamefont
  {Vidal}}\ and\ \bibinfo {author} {\bibfnamefont {R.~F.}\ \bibnamefont
  {Werner}},\ }\bibfield  {title} {\enquote {\bibinfo {title} {Computable
  measure of entanglement},}\ }\href {\doibase 10.1103/PhysRevA.65.032314}
  {\bibfield  {journal} {\bibinfo  {journal} {Phys. Rev. A}\ }\textbf {\bibinfo
  {volume} {65}},\ \bibinfo {pages} {032314} (\bibinfo {year}
  {2002})}\BibitemShut {NoStop}%
\bibitem [{\citenamefont {Zhao}\ \emph {et~al.}(2016)\citenamefont {Zhao},
  \citenamefont {Yuan},\ and\ \citenamefont {Ma}}]{PhysRevA.94.012343}%
  \BibitemOpen
  \bibfield  {author} {\bibinfo {author} {\bibfnamefont {Qi}~\bibnamefont
  {Zhao}}, \bibinfo {author} {\bibfnamefont {Xiao}\ \bibnamefont {Yuan}}, \
  and\ \bibinfo {author} {\bibfnamefont {Xiongfeng}\ \bibnamefont {Ma}},\
  }\bibfield  {title} {\enquote {\bibinfo {title} {Efficient
  measurement-device-independent detection of multipartite entanglement
  structure},}\ }\href {\doibase 10.1103/PhysRevA.94.012343} {\bibfield
  {journal} {\bibinfo  {journal} {Phys. Rev. A}\ }\textbf {\bibinfo {volume}
  {94}},\ \bibinfo {pages} {012343} (\bibinfo {year} {2016})}\BibitemShut
  {NoStop}%
\bibitem [{\citenamefont {Cruz}\ \emph {et~al.}(2019)\citenamefont {Cruz},
  \citenamefont {Fournier}, \citenamefont {Gremion}, \citenamefont {Jeannerot},
  \citenamefont {Komagata}, \citenamefont {Tosic}, \citenamefont
  {Thiesbrummel}, \citenamefont {Chan}, \citenamefont {Macris}, \citenamefont
  {Dupertuis},\ and\ \citenamefont {Javerzac-Galy}}]{Cruz2019}%
  \BibitemOpen
  \bibfield  {author} {\bibinfo {author} {\bibfnamefont {Diogo}\ \bibnamefont
  {Cruz}}, \bibinfo {author} {\bibfnamefont {Romain}\ \bibnamefont {Fournier}},
  \bibinfo {author} {\bibfnamefont {Fabien}\ \bibnamefont {Gremion}}, \bibinfo
  {author} {\bibfnamefont {Alix}\ \bibnamefont {Jeannerot}}, \bibinfo {author}
  {\bibfnamefont {Kenichi}\ \bibnamefont {Komagata}}, \bibinfo {author}
  {\bibfnamefont {Tara}\ \bibnamefont {Tosic}}, \bibinfo {author}
  {\bibfnamefont {Jarla}\ \bibnamefont {Thiesbrummel}}, \bibinfo {author}
  {\bibfnamefont {Chun~Lam}\ \bibnamefont {Chan}}, \bibinfo {author}
  {\bibfnamefont {Nicolas}\ \bibnamefont {Macris}}, \bibinfo {author}
  {\bibfnamefont {Marc-André}\ \bibnamefont {Dupertuis}}, \ and\ \bibinfo
  {author} {\bibfnamefont {Clément}\ \bibnamefont {Javerzac-Galy}},\
  }\bibfield  {title} {\enquote {\bibinfo {title} {Efficient quantum algorithms
  for ghz and w states, and implementation on the ibm quantum computer},}\
  }\href {\doibase https://doi.org/10.1002/qute.201900015} {\bibfield
  {journal} {\bibinfo  {journal} {Advanced Quantum Technologies}\ }\textbf
  {\bibinfo {volume} {2}},\ \bibinfo {pages} {1900015} (\bibinfo {year}
  {2019})}\BibitemShut {NoStop}%
\bibitem [{\citenamefont {García-Pérez}\ \emph {et~al.}(2021)\citenamefont
  {García-Pérez}, \citenamefont {Rossi}, \citenamefont {Sokolov},
  \citenamefont {Tacchino}, \citenamefont {Barkoutsos}, \citenamefont
  {Mazzola}, \citenamefont {Tavernelli},\ and\ \citenamefont
  {Maniscalco}}]{garciaperez2021learning}%
  \BibitemOpen
  \bibfield  {author} {\bibinfo {author} {\bibfnamefont {Guillermo}\
  \bibnamefont {García-Pérez}}, \bibinfo {author} {\bibfnamefont {Matteo
  A.~C.}\ \bibnamefont {Rossi}}, \bibinfo {author} {\bibfnamefont {Boris}\
  \bibnamefont {Sokolov}}, \bibinfo {author} {\bibfnamefont {Francesco}\
  \bibnamefont {Tacchino}}, \bibinfo {author} {\bibfnamefont {Panagiotis~Kl.}\
  \bibnamefont {Barkoutsos}}, \bibinfo {author} {\bibfnamefont {Guglielmo}\
  \bibnamefont {Mazzola}}, \bibinfo {author} {\bibfnamefont {Ivano}\
  \bibnamefont {Tavernelli}}, \ and\ \bibinfo {author} {\bibfnamefont
  {Sabrina}\ \bibnamefont {Maniscalco}},\ }\href@noop {} {\enquote {\bibinfo
  {title} {{Learning to measure: adaptive informationally complete POVMs for
  near-term quantum algorithms}},}\ } (\bibinfo {year} {2021}),\ \Eprint
  {http://arxiv.org/abs/2104.00569} {arXiv:2104.00569 [quant-ph]} \BibitemShut
  {NoStop}%
\bibitem [{\citenamefont {Garc\'{i}a-P\'{e}rez}\ \emph
  {et~al.}(2021)\citenamefont {Garc\'{i}a-P\'{e}rez}, \citenamefont {Kerppo},
  \citenamefont {Rossi},\ and\ \citenamefont {Maniscalco}}]{github}%
  \BibitemOpen
  \bibfield  {author} {\bibinfo {author} {\bibfnamefont {Guillermo}\
  \bibnamefont {Garc\'{i}a-P\'{e}rez}}, \bibinfo {author} {\bibfnamefont
  {Oskari}\ \bibnamefont {Kerppo}}, \bibinfo {author} {\bibfnamefont {Matteo
  A.~C.}\ \bibnamefont {Rossi}}, \ and\ \bibinfo {author} {\bibfnamefont
  {Sabrina}\ \bibnamefont {Maniscalco}},\ }\href
  {https://github.com/matteoacrossi/non-separability-multipartite-entanglement}
  {\enquote {\bibinfo {title}
  {github.com/matteoacrossi/non-separability-multipartite-entanglement},}\ }
  (\bibinfo {year} {2021})\BibitemShut {NoStop}%
\end{thebibliography}%

\onecolumngrid

\clearpage

\appendix
\setcounter{figure}{0}
\setcounter{page}{1}
\renewcommand{\thefigure}{S\arabic{figure}}
\renewcommand{\theequation}{S\arabic{equation}}

\begin{center}
	\textbf{\large Supplementary Material\\
	``Experimentally accessible non-separability criteria\\for multipartite entanglement structure detection''}
\end{center}

\begin{figure*}[h]
    \centering
    \includegraphics[width=.69\linewidth]{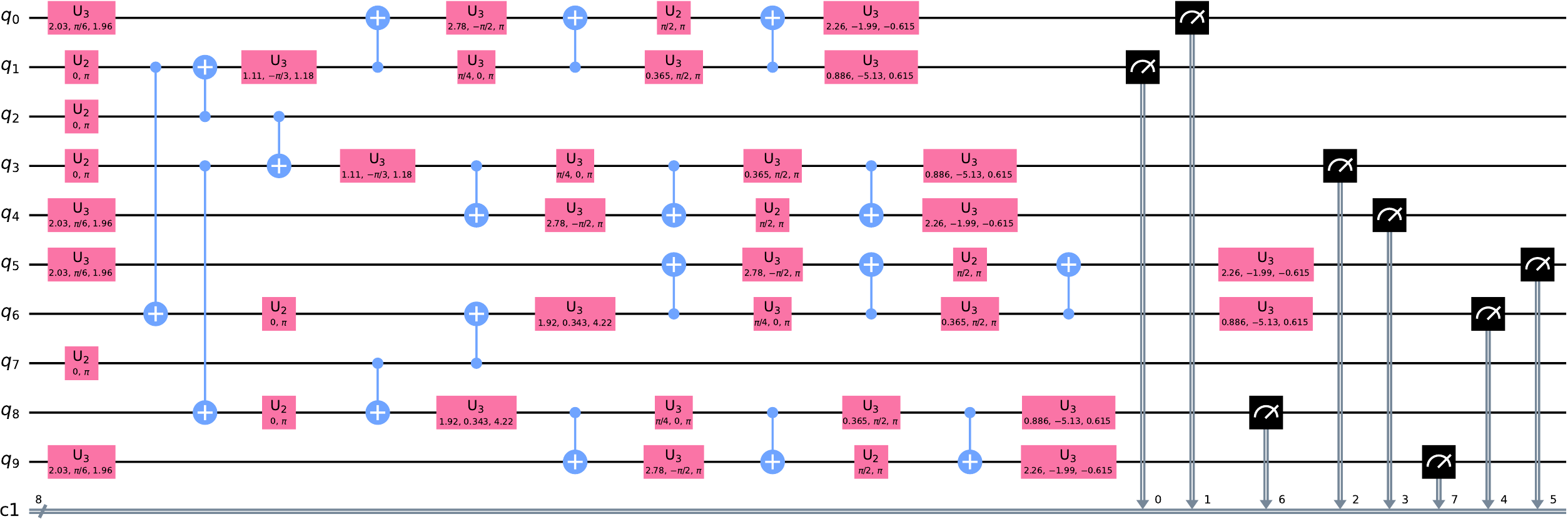}\hfill\includegraphics[width=.25\linewidth]{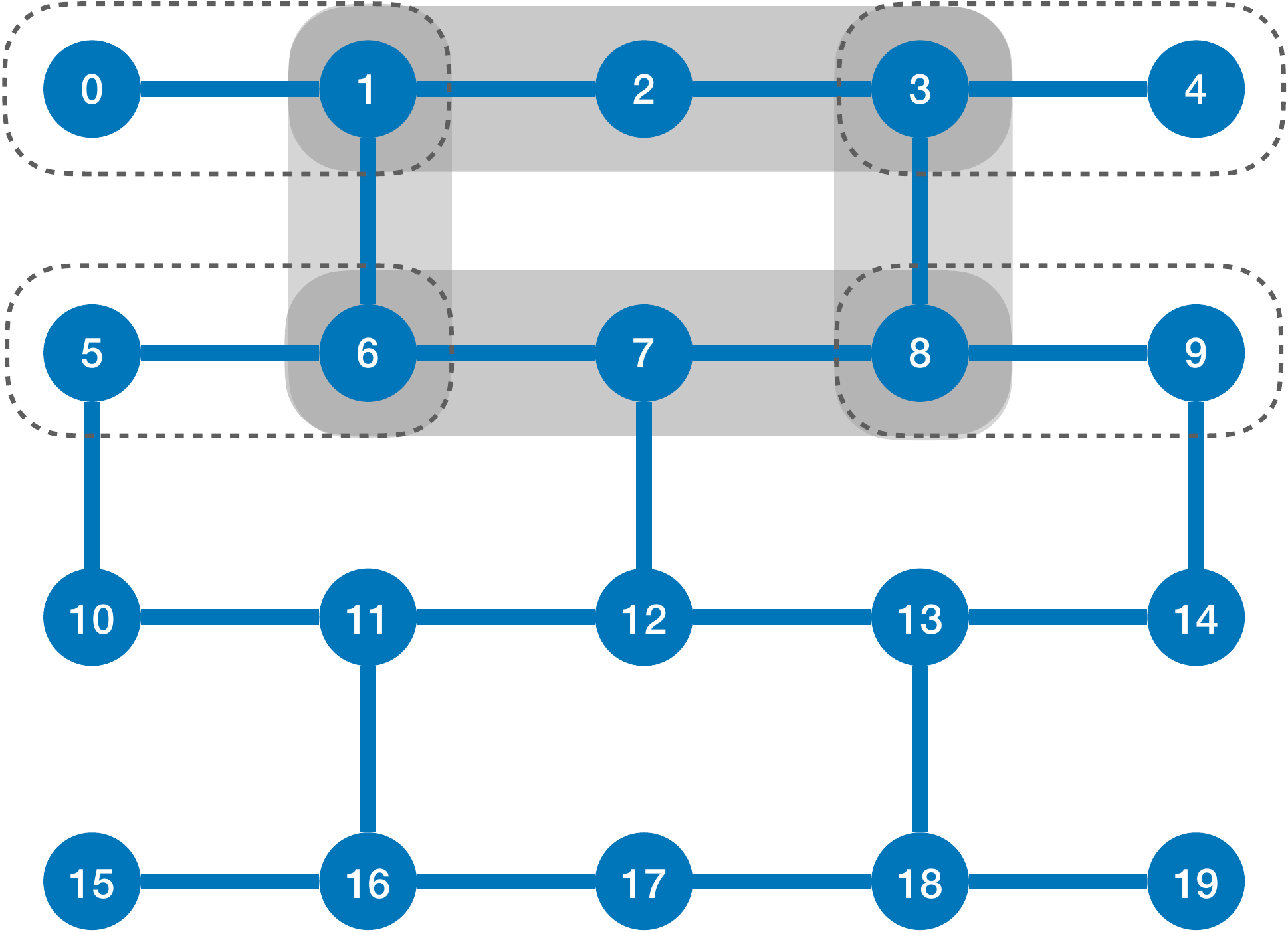}
    \caption{Left: circuit implementation of the Smolin state in terms of the native gates of the \texttt{ibmq\_singapore} device. Right: layout of the device, highlighted in grey the qubits used for the system ($q_1$, $q_3$, $q_6$ and $q_8$ and the ancillae $q_2$ and $q_7$). The dashed lines show the pairing between system qubits and auxiliary qubits used for the implementation of the POVMs.}
    \label{fig:smolin_circuit}
\end{figure*}

\begin{figure*}[h]
    \centering
    \includegraphics[width=\linewidth]{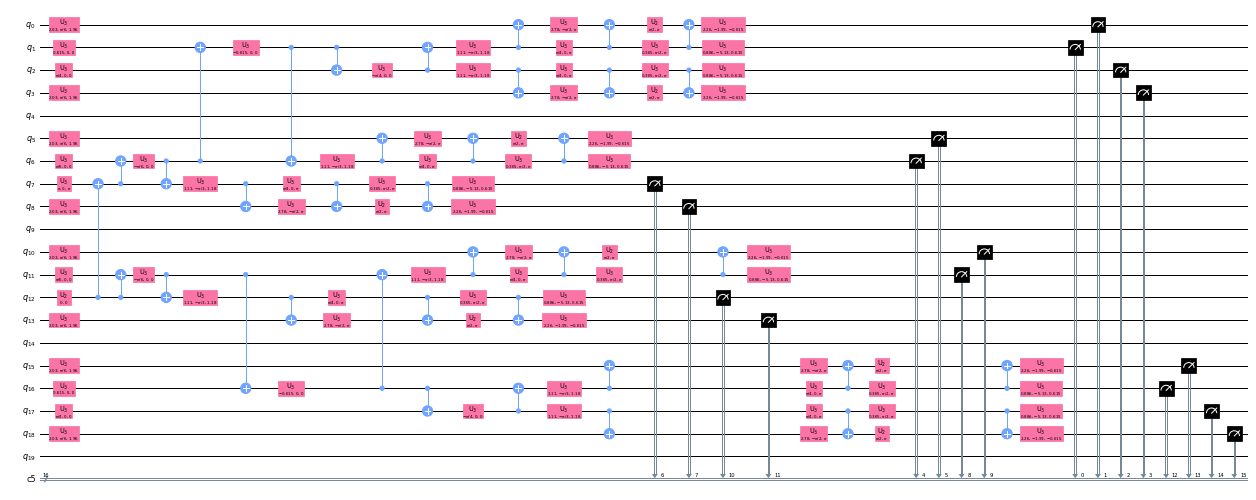}
    \caption{Circuit implementations of the W state in terms of the native gates of the \texttt{ibmq\_singapore} device.}
    \label{fig:w_circuit}
\end{figure*}

\begin{figure*}[h]
\includegraphics[width=.4\columnwidth]{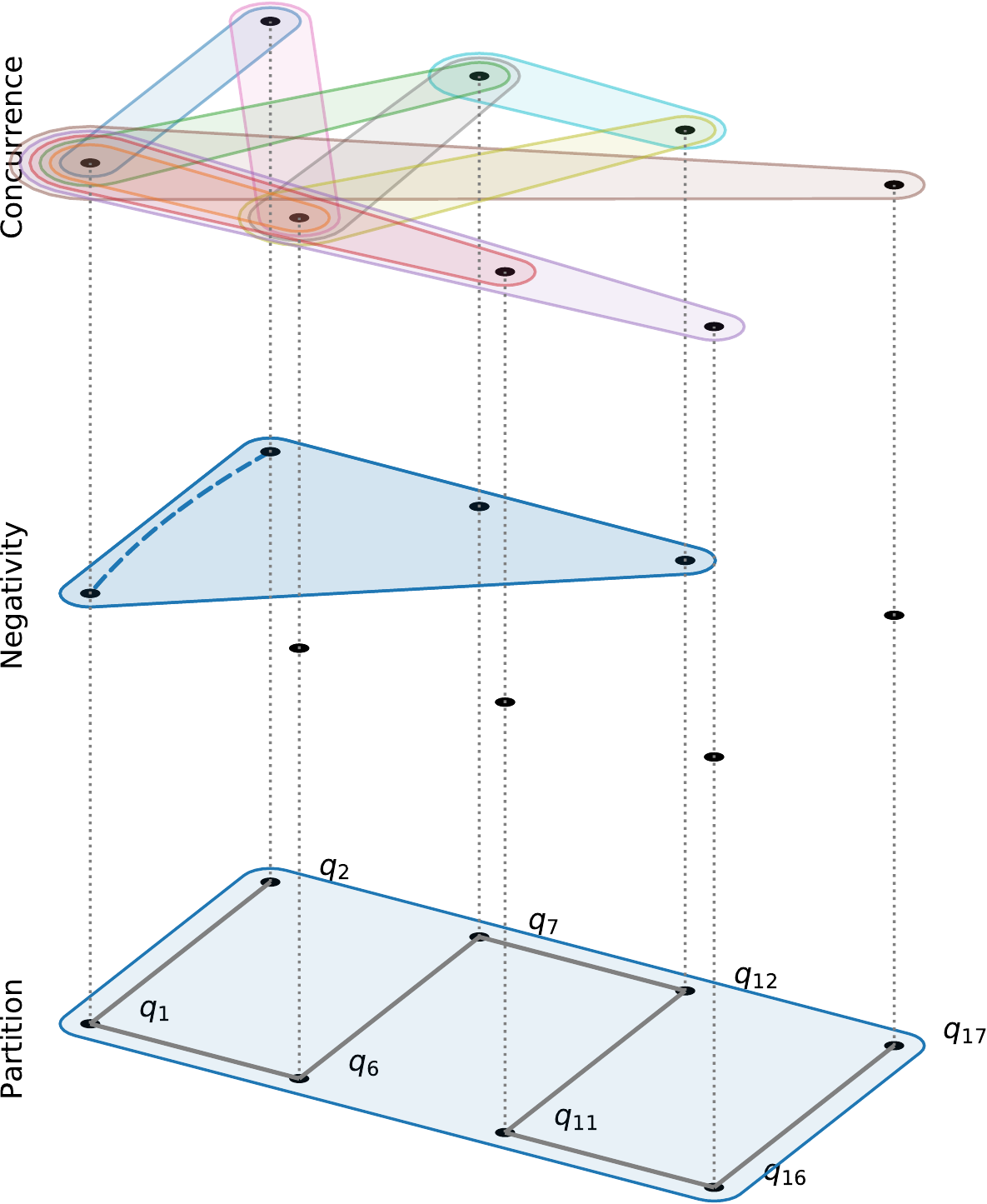}
\includegraphics[width=.4\columnwidth]{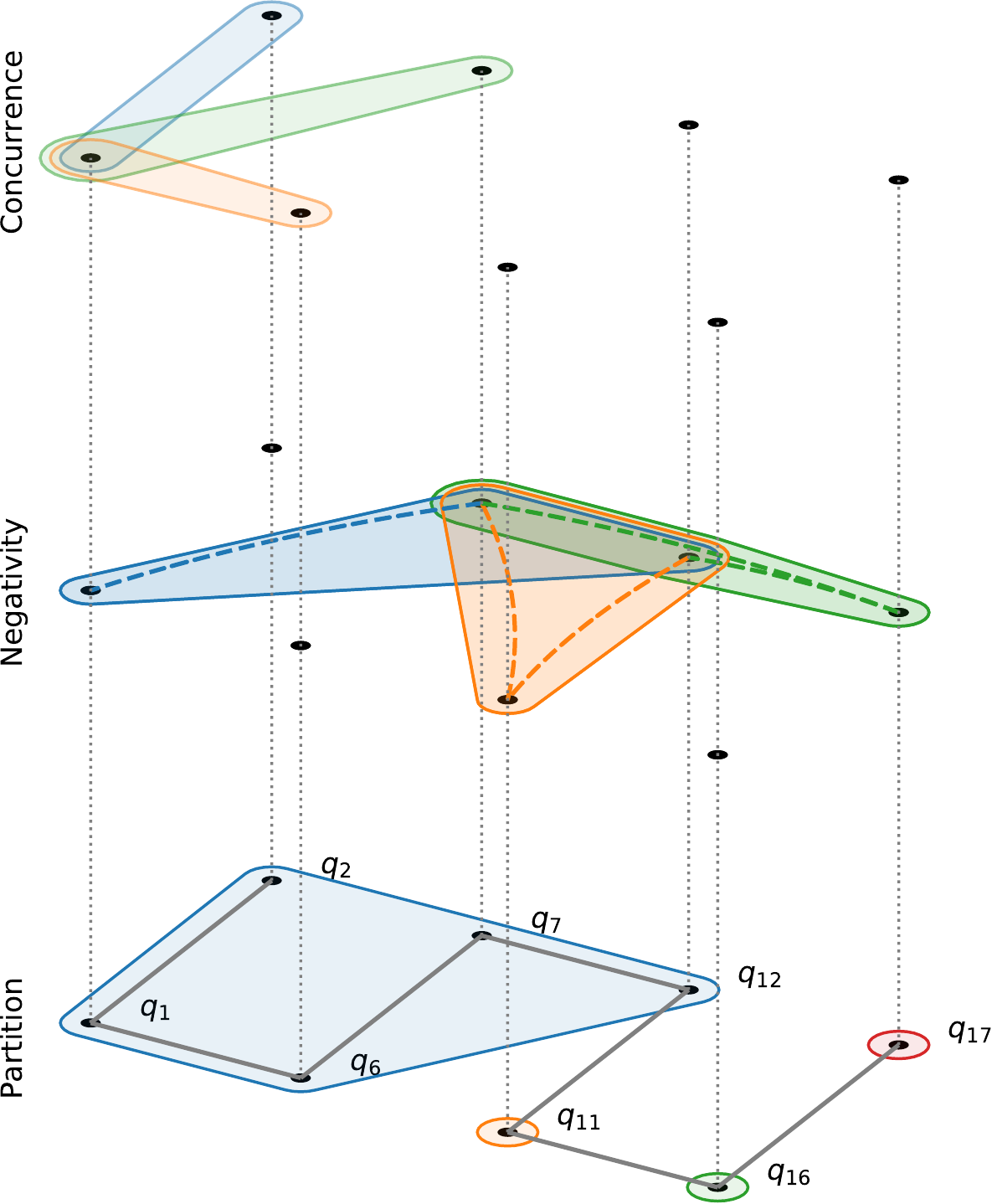}
\vspace{1cm}
\includegraphics[width=.4\columnwidth]{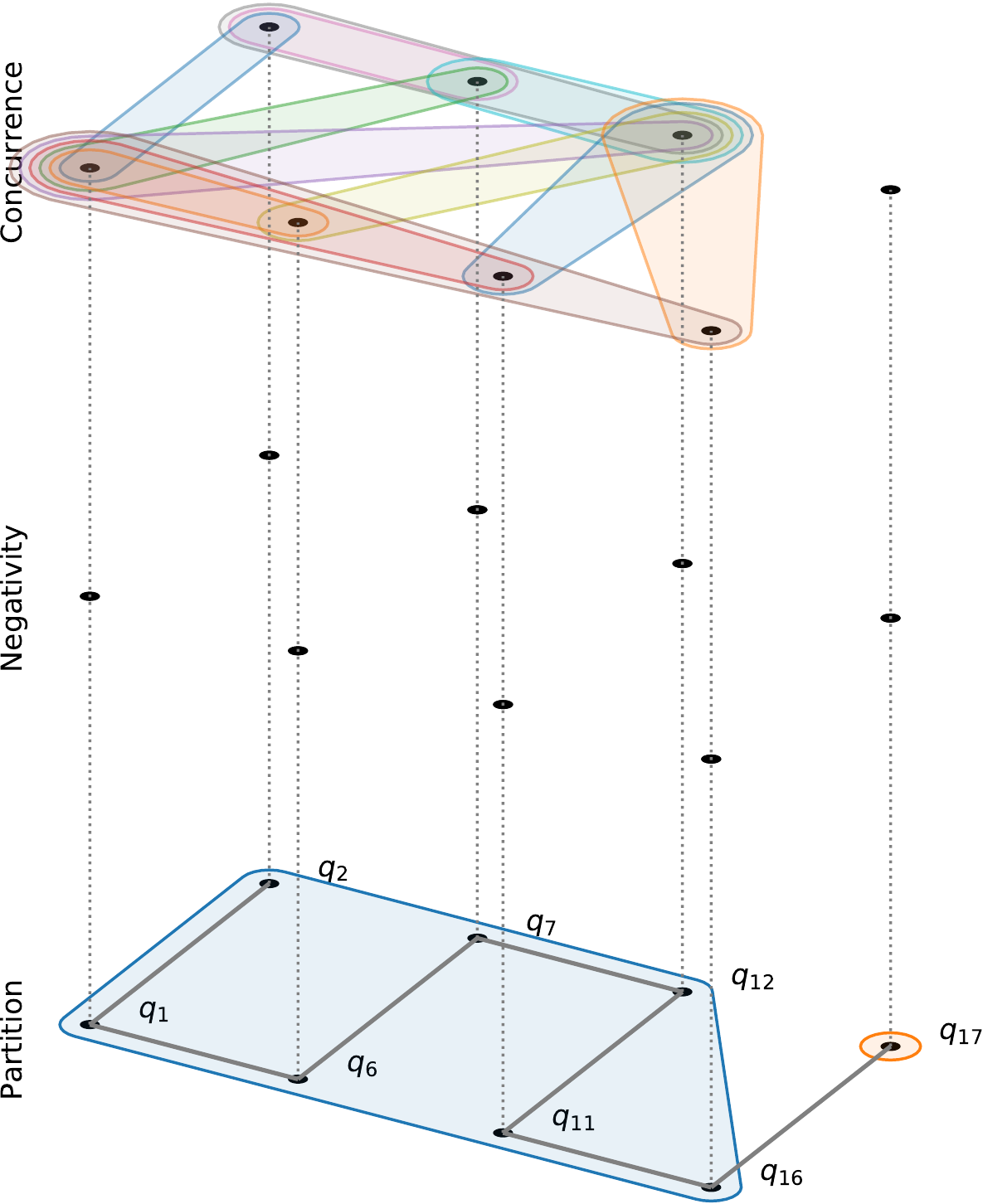}
\includegraphics[width=.4\columnwidth]{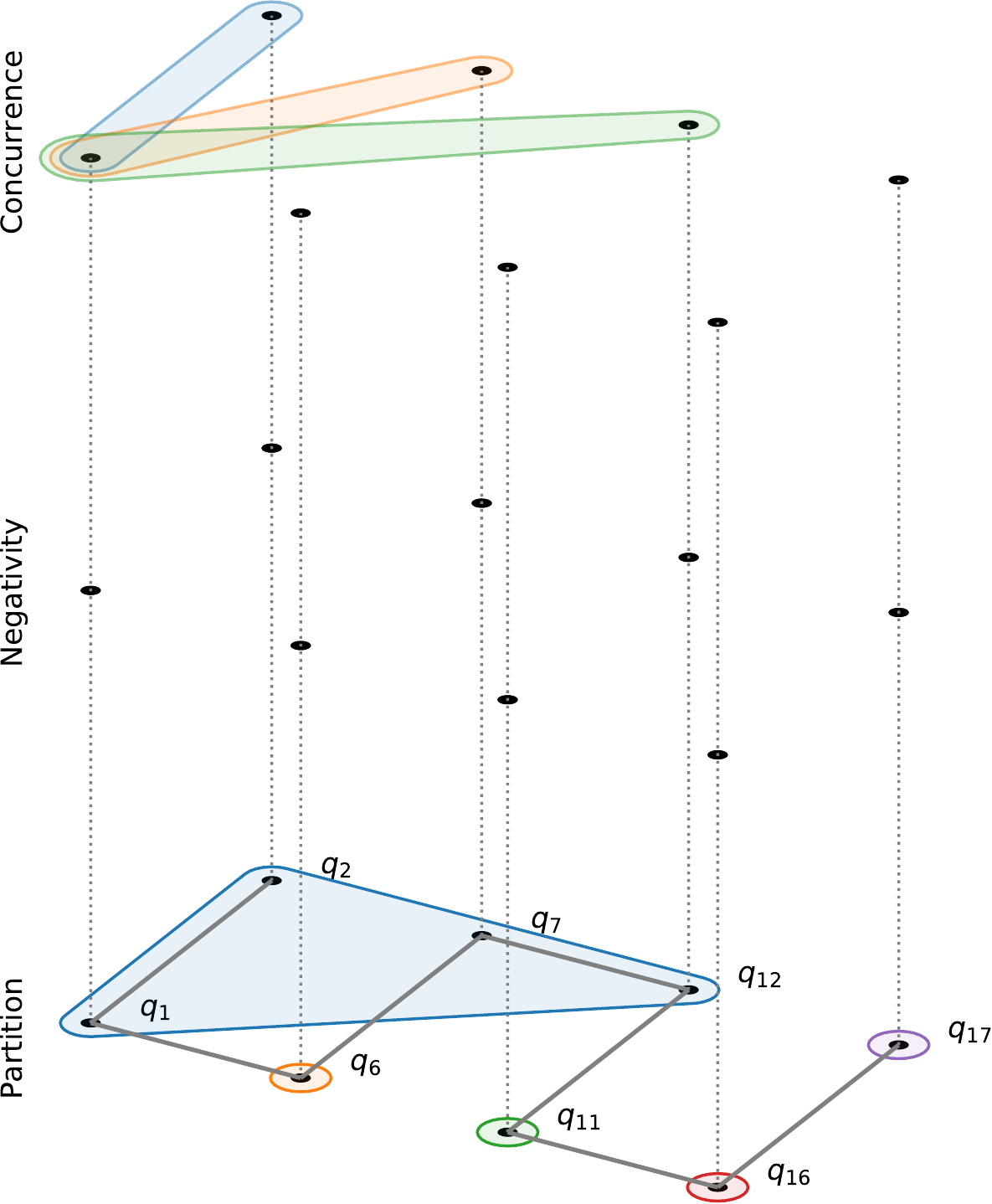}
\caption{Multiplex plots of the experimentally reconstructed multipartite entanglement structure of the W state on \texttt{ibmq\_singapore} using qubits $q_1$, $q_2$, $q_6$, $q_7$, $q_{11}$, $q_{12}$, $q_{16}$, and $q_{17}$. Each figure corresponds to a different experimental run.}
\label{fig:w_multiplexes_left}
\end{figure*}

\begin{figure*}[h]
\hfill\raisebox{\height}{\includegraphics[width=.25\columnwidth]{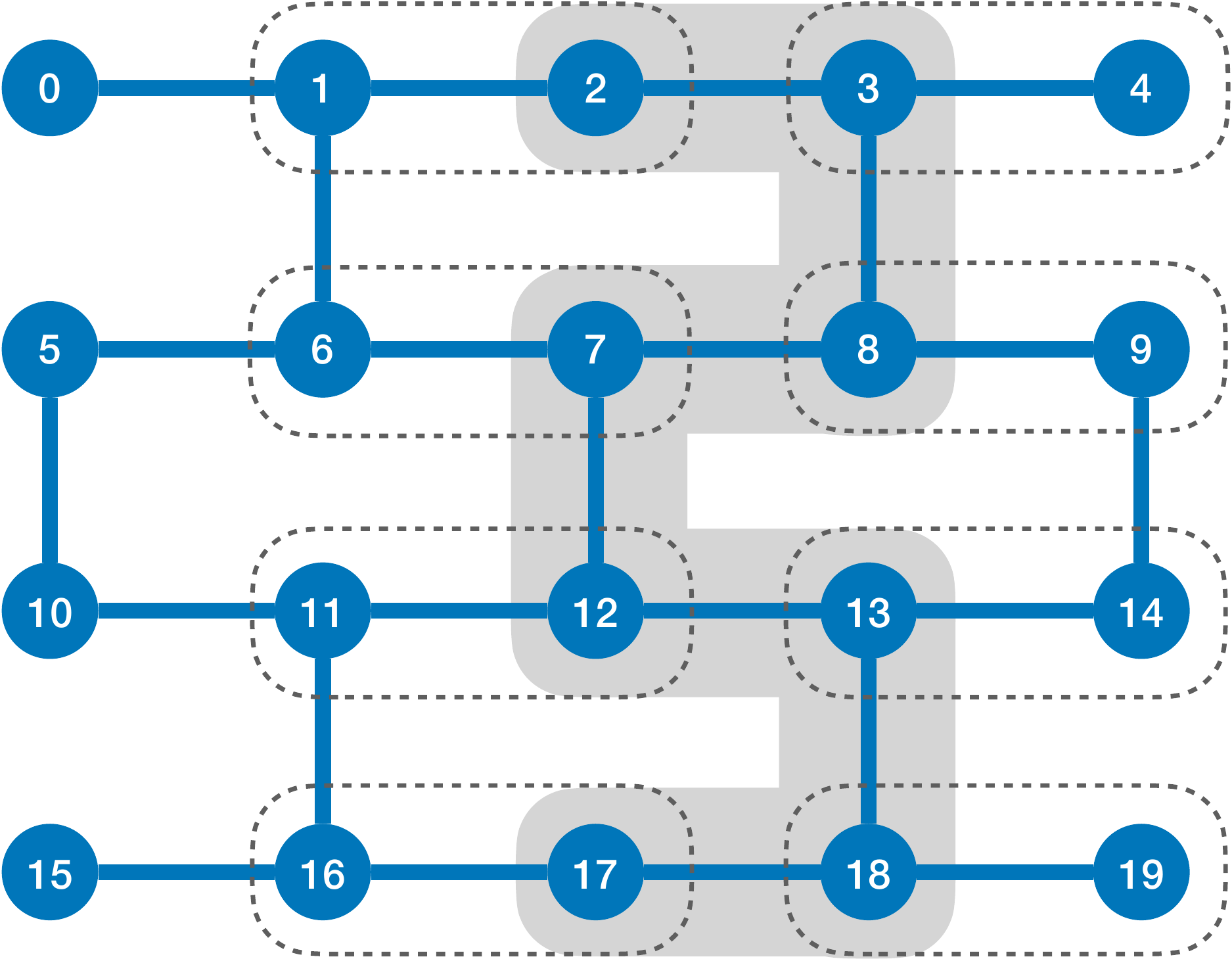}}\hfill
\includegraphics[width=.32\columnwidth]{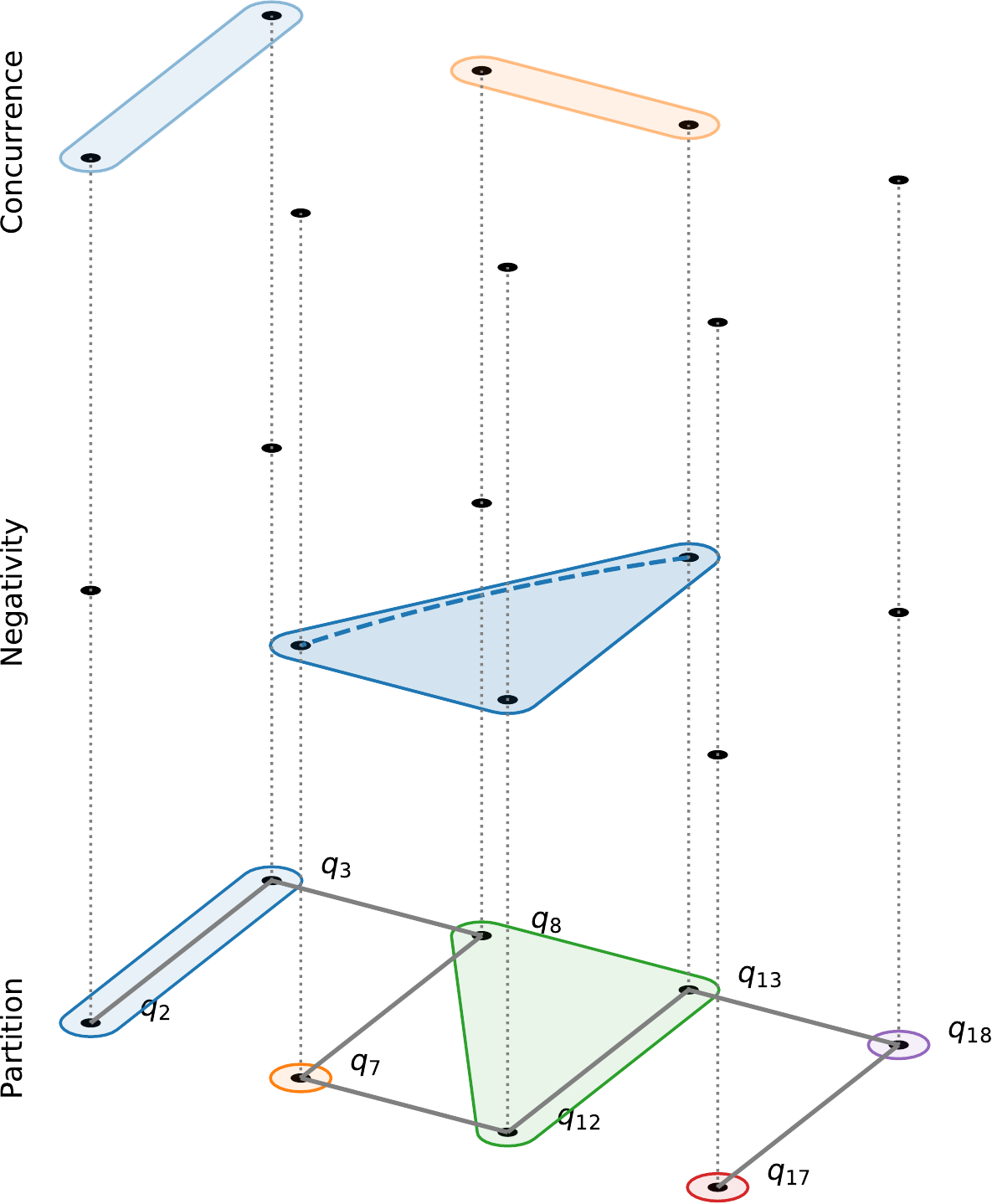}
\includegraphics[width=.32\columnwidth]{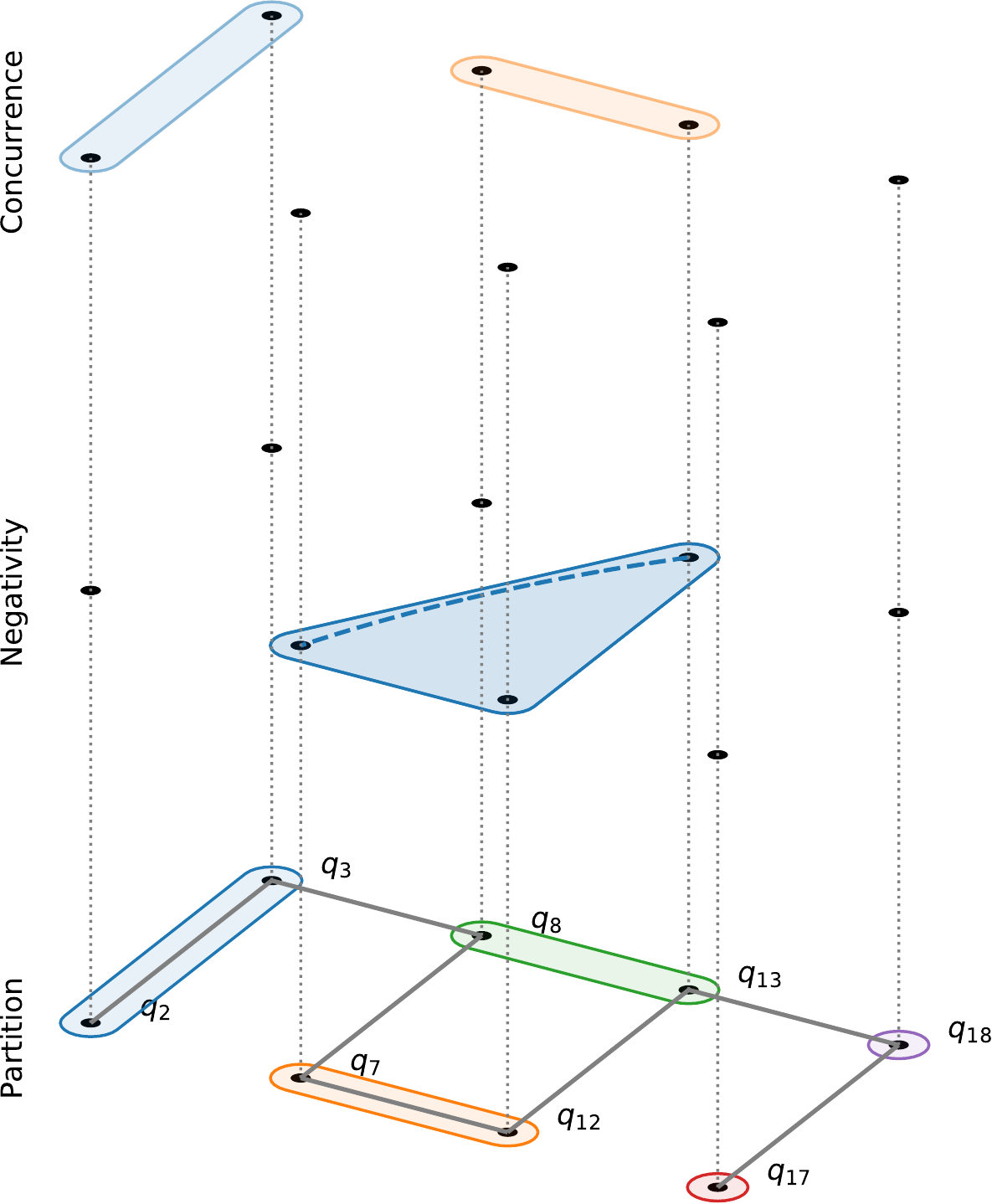}
\vspace{1cm}

\includegraphics[width=.32\columnwidth]{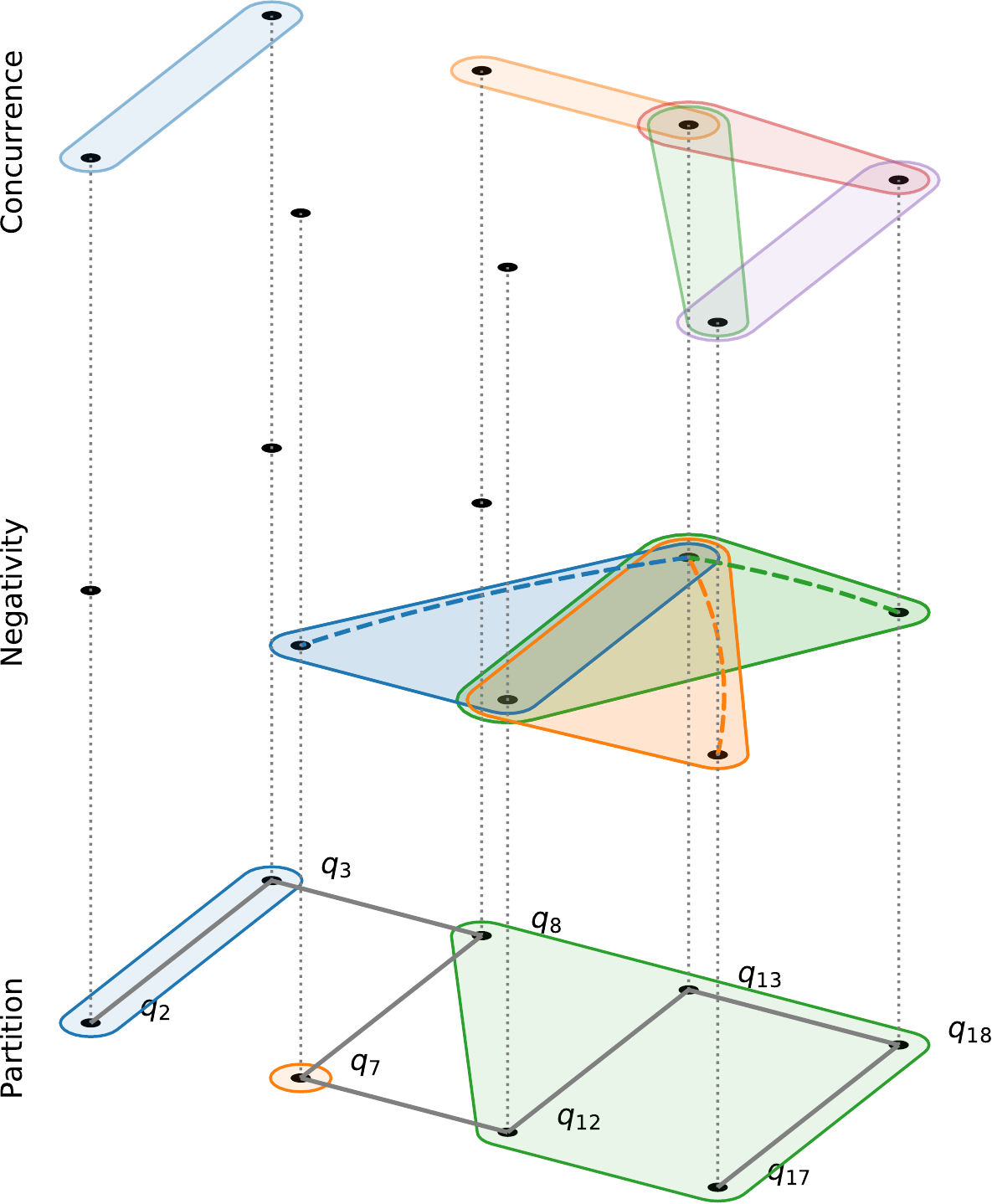}
\includegraphics[width=.32\columnwidth]{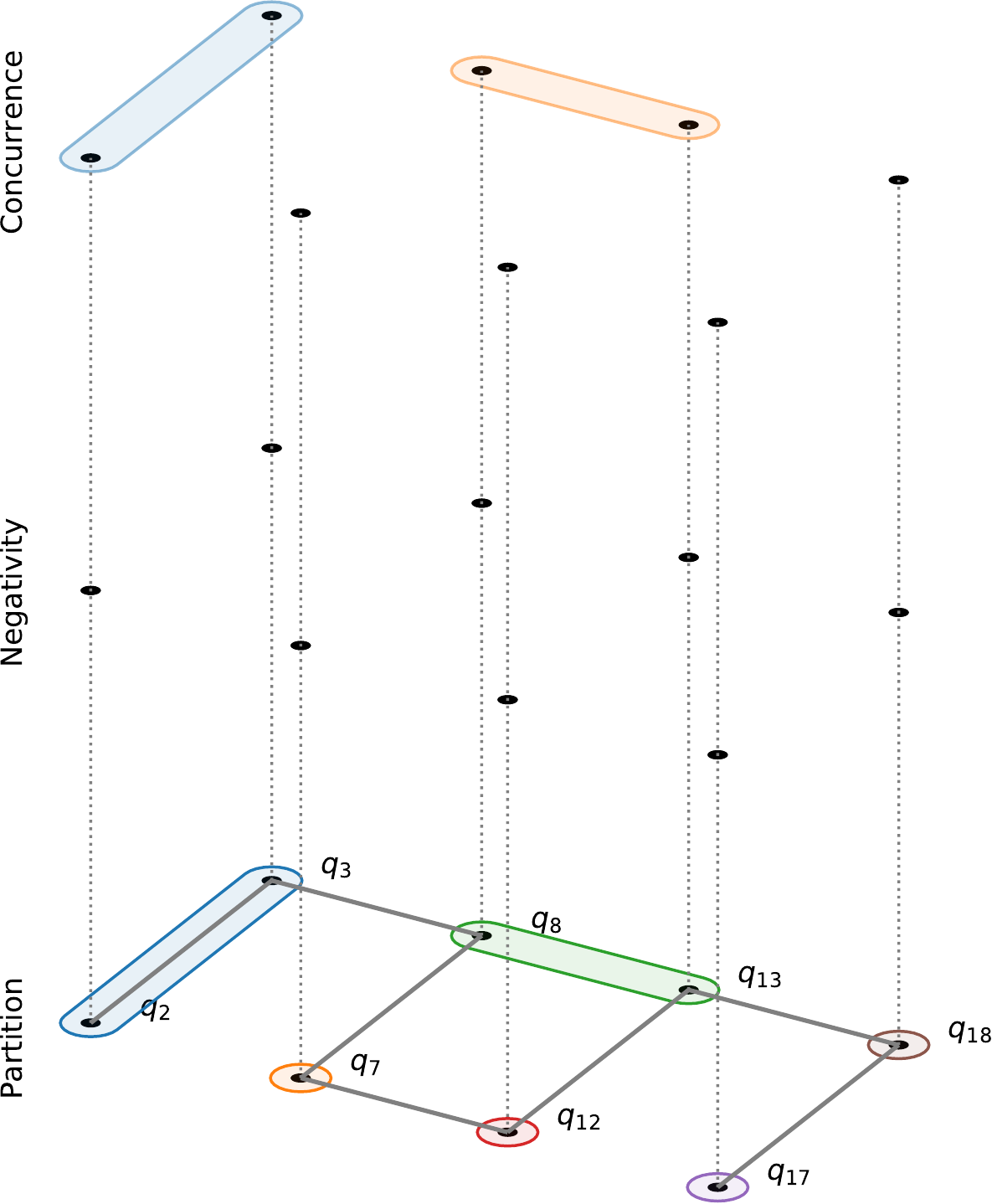}
\includegraphics[width=.32\columnwidth]{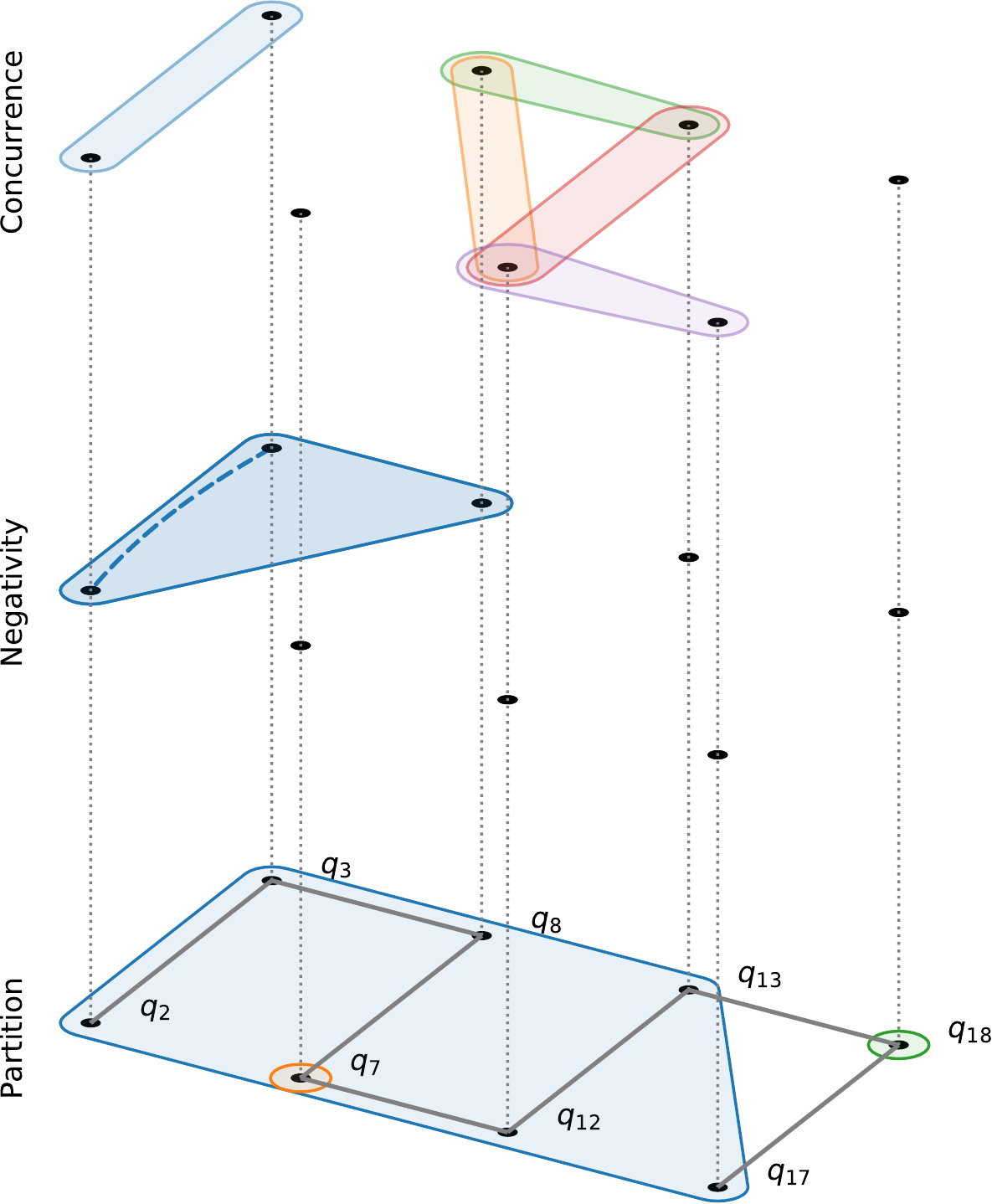}
\caption{Multiplex plots of the experimentally reconstructed multipartite entanglement structure of the W state on \texttt{ibmq\_singapore} using qubits $q_2$, $q_3$, $q_7$, $q_8$, $q_{12}$, $q_{13}$, $q_{17}$, and $q_{18}$, as indicated in the schematic representation of the device. The figures correspond to four different experimental runs. Notice that the top two figures depict the two minimal partitions for the same realisation.}
\label{fig:w_multiplexes_right}
\end{figure*}

\end{document}